\documentclass[1p]{elsarticle}

\usepackage{lineno,hyperref}
\hypersetup{colorlinks = true, allcolors = cyan}

\makeatletter
\def\ps@pprintTitle{%
   \let\@oddhead\@empty
   \let\@evenhead\@empty
   \let\@oddfoot\@empty
   \let\@evenfoot\@oddfoot
}
\makeatother

\usepackage{nostredef}
\usepackage{cleveref}
\creflabelformat{subsection}{section}

\modulolinenumbers[5]

\begin{document}

\begin{frontmatter}

\title{Computing multiparameter persistent homology through a discrete Morse-based approach}

\author[first-affiliation]{Sara Scaramuccia}
\ead{sara.scaramuccia@dibris.unige.it}
\author[second-affiliation]{Federico Iuricich\corref{mycorrespondingauthor}}
\ead{fiurici@clemson.edu}
\author[third-affiliation]{Leila De Floriani}
\ead{deflo@umiacs.umd.edu}
\author[fourth-affiliation]{Claudia Landi}
\ead{claudia.landi@unimore.it}

\cortext[mycorrespondingauthor]{Corresponding author}

\address[first-affiliation]{University of Genova, Genova, Italy}
\address[second-affiliation]{Clemson University, Clemson (SC), USA}
\address[third-affiliation]{University of Maryland, College Park (MD), USA }
\address[fourth-affiliation]{University of Modena and Reggio Emilia, Italy}

\begin{abstract}
%
%
%
{\em Persistent Homology} (PH) allows tracking homology features like loops, holes and their higher-dimensional analogs, along with a single-parameter family of nested spaces.
Currently, computing descriptors for complex data characterized by multiple functions is becoming a major challenging task in several applications, including physics, chemistry, medicine, geography, etc.
{\em Multiparameter Persistent Homology} (MPH) generalizes persistent homology opening to the exploration and analysis of shapes endowed with multiple filtering functions. Still, computational constraints prevent MPH to be feasible over real-sized data.
In this paper, we consider {\em discrete Morse Theory} \cite{Forman1998} as a tool to simplify the computation of MPH on a multiparameter dataset. We propose a new algorithm, well suited for parallel and distributed implementations and we provide the first evaluation of the impact on MPH computations of a preprocessing approach.
\end{abstract}


\end{frontmatter}

\section{Introduction}
\label{sec:intro}



In recent years, the increasing amount of data has led to the improvement and development of information handling techniques.
  The basic goal of Topological Data Analysis (TDA) is to retrieve and organize qualitative information about data.

{\em Homology} is one of the most relevant invariants studied in TDA but has the drawback of being scarcely descriptive.

{\em Persistent Homology (PH)} allows for a multiresolution analysis of homologies by means of a {\em filtration}.
  It is used in data analysis to study evolutions of qualitative features of data and it is appreciated for its computability, robustness to noise, and dimension independence.
So far, many optimization methods in computing PH have been proposed.
Those more tightly related to this paper refer to another relevant tool for TDA, namely {\em discrete Morse theory}~\cite{Forman1998}.
In this case, discrete Morse theory provides an important preprocessing tool for homology computations by defining a {\em discrete gradient field} (also called discrete gradient) over the input datum.
This allows reducing the size of the input space to the critical parts, generally few, with respect to the retrieved discrete gradient.
The discrete gradient can be also built so that to preserve the filtration structure thus enhancing PH computations.
Although other PH optimizations outperform this Morse-based preprocessing,  this no longer applies to the generalization of PH called {\em multiparameter persistent homology} (MPH).

MPH is an extension of the PH theory motivated by the fact that data analysis and comparisons often involve the examination of properties that are naturally described by multiple parameters, for instance in computer vision with respect to photometric properties. Alternatively, for point cloud data, several criteria might be chosen in order to filter the input datum for investigating both the domain itself under several criteria at once and the explanatory power of the different criteria over the same domain.

The entire information provided by MPH is captured by the {\em persistence module}.
Alternatively, the {\em persistence space} summarizes the MPH information into a collection of PH descriptors.

All available MPH methods suffer from high computational costs and scalability problems.
This prevents them to be feasible over real-sized data sets.
A Morse-based preprocessing solution, generalized to the multiparameter case, has been proposed in  \cite{Allili2017dgci}.
This can have, in theory, a valuable impact on MPH computations. However, that preprocessing still presents limitations in scalability with real data.

We propose here the first algorithm capable of computing a discrete gradient on real-world data. Our approach is easy-to-use and well suited for parallel and distributed implementations. We consider the applicative domain where the obtained representation can be successfully adopted, namely, for reducing the complexity of computing MPH.

Our contributions is composed of:
\begin{itemize}
	\item {\textbf{a new algorithm }} for retrieving a discrete gradient which preserves MPH and suitable for real-sized data sets;
	\item {\textbf{a comparison}} of the scalability of our proposed algorithm with respect to the equivalent method in state of the art;
  \item {\textbf{an evaluation }} of the advantages obtained by using our proposed preprocessing in persistence module and persistence space computation.
\end{itemize}

The remainder of this paper is organized as follows.
In Section \ref{sec:background}, we introduce the notions at the basis of our work. Related work is reviewed in Section \ref{sec:related}.
The new preprocessing algorithm is described in Section \ref{sec:algo} where we also present a detailed analysis of complexity.
In Section \ref{sec:correctness}, we present the proof of correctness and a theoretical and experimental comparison, of our approach, with the one presented in \cite{Allili2017dgci}.
The results of computing MPH with our approach are discussed in Section \ref{sec:exp}.
In Section \ref{sec:conclusions}, we draw concluding remarks and we discuss future developments.

\section{Background}
\label{sec:background}

\subsection{Simplicial complexes}
\label{sec:input/output}
A simplicial complex is a discrete topological structure introduced to formalize the input of our algorithm.
A $k$-dimensional simplex $\sigma$, or $k$-simplex for short, is the convex hull of $k+1$ affinely independent points. Often, we will write $\s^k$ to shortly mean a $k$-simplex.
A {\em face} $\tau$ of $\sigma$ is the convex hull of any subset of points generating $\sigma$. The partial order relation ``$\tau$ is face of $\s$'' is denoted $\tau\ll\s$.
If the dimensions of $\tau$ and $\sigma$ differ by one we call $\tau$ a {\em facet} of $\s$ and denote it by $\tau<\sigma$.
Dually, $\sigma$ is a {\em coface} of $\tau$ and a {\em cofacet} when the two dimensions differ by one (respectively indicated $\s\gg\tau$ and $\sigma>\tau$).
\begin{deff}[Simplicial complex]
	A {\em simplicial complex} $\simplicial$ is a finite collection of simplices such that:
	\begin{itemize}
		\item every face of a simplex in $\simplicial$ is also in $\simplicial$
		\item {\em intersection property: }the intersection of any two simplices in $\simplicial$ is either empty or a single simplex in $\simplicial$.
	\end{itemize}
	\label{def:simplicial}
\end{deff}
We will denote by $\simplicial_k$ the set of $k$-simplices in $\simplicial$. An element in $\simplicial_0$ is also called a {\em vertex}.
A simplicial complex $\simplicial$ has dimension $d$ ($d$-complex for short) if the maximum of the dimensions of its simplices is $d$.
%
The {\em boundary} of a simplex $\s$ in $\simplicial$ is the set of all faces of $\s$ in $\simplicial$.
The {\em coboundary} (or the {\em star}) of $\s$ in $\simplicial$ is the set of cofaces of $\s$ in $\simplicial$.

\subsection{Discrete Morse theory}

The algorithm proposed in this paper retrieves a combinatorial object called  {\em discrete gradient} over the domain $ \simplicial$.
The relevance of this output has to be seen within the framework of Forman's discrete Morse Theory~\cite{Forman1998}.
%
In discrete Morse Theory, a {\em (discrete) vector} is a pair of simplices $(\sigma,\tau)$ such that $\sigma<\tau$.
A {\em discrete vector field} is any collection of vectors $V$ such that each simplex is component of at most one vector in $V$.
%
A $V${\em -path} is a sequence of vectors $(\sigma_i,\tau_i)$ belonging to $V$, for $i=0,\dots,r$, such that, for all indexes $0\leq i\leq r-1$, $\sigma_{i+1}<\tau_i$ and $\sigma_i\neq\sigma_{i+1}$.
A $V$-path is said to be {\em closed} if $\sigma_0=\sigma_r$ and {\em trivial} if $r=0$.
 We say that a $V$-path $\{ (\sigma_0^{k},\tau_0^{k+1}),\dots, (\sigma_l^{k},\tau_l^{k+1}) \}$ is {\em from $\s^k$ to $\bs^k$}, if $\sigma_0^{k}= \s^k$ and $\tau_l^{k+1}>\bs^k$. If $\s^k = \bs^k$ the empty set is a valid $V$-path from $\s^k$ to itself.
\begin{deff}[discrete gradient]
	A discrete vector field $V$ is a discrete {\em gradient} if all of its closed $V$-paths are trivial.
	\label{def:gradient}
\end{deff}

Simplices that do not belong to any vector are said {\em critical}. Moreover, instead of critical simplices, we use the term critical {\em cell} with dimension equal to the simplex dimension.

The set of critical cells $\critical(V)$ is called a {\em critical set over } $\complex$.
From now on, we wright $\critical$ instead of $\critical(V)$ when the discrete gradient $V$ is understood.
Given a discrete gradient $V$, a {\em separatrix from a critical cell $\tau^{k+1}$ to a critical cell $\s^k$} is a $V$-path from any face of $\tau^{k+1}$ to $\s^k$.

\subsection{Homology}
\label{sec:hom}
%
Intuitively, the homology of a simplicial complex $\simplicial$ detects independent $k$-dimensional cycles of $\simplicial$, i.e., different connected components (0-cycles), tunnels (1-cycles), voids (2-cycles), and so on.
More precisely, the incidence relations among simplices in a simplicial complex are combinatorially translated into algebraic terms and, the intuition of a loop is captured by linear combinations of simplices called {\em chains}.
We focus on the case of linear combinations over $\F_2$ the field with the two elements $0$ and $1$.
Incidence relations are translated into algebraic terms by means of a suitable incidence function.

The {\em simplicial incidence function} $\sif:\simplicial\times \simplicial\longrightarrow \F_2$ associated with a simplicial complex $\simplicial$ is defined by
\begin{align*}
\sif(\tau,\s)=1
\qquad&\leftrightarrow \qquad
\tau>\s
\\
\sif(\tau,\s)=0
\qquad&\leftrightarrow \qquad
\mbox{ otherwise}
\end{align*}

It is not necessary to have a simplicial complex to construct a chain complex, and thus, to get the homology associated with it.
Indeed, in the case of a discrete gradient $V$ over $\simplicial$, we define 
a {\em separatrix from a critical cell $\tau^{k+1}$ to a critical cell $\s^k$} to be a $V$-path from any face of $\tau^{k+1}$ to $\s^k$.
Then, we can translate incidence relations among critical cells in $\critical$ into algebraic terms by means of the {\em critical incidence function} $\mif:\critical\times \critical\longrightarrow \F_2$ defined by
\begin{align*}
\mif(\tau,\s)=1
\qquad&\leftrightarrow\qquad
\mbox{ the number of different separatrices from } \tau \mbox{ to } \s \mbox{ is odd}
\\
\mif(\tau,\s)=0
\qquad&\leftrightarrow\qquad
\mbox{ the number of different separatrices from } \tau \mbox{ to } \s  \mbox{ is even}
\end{align*}
We remark that the definition of critical incidence function in this form comes from Theorem 8.10 in \cite{Forman1998} in the particular case of $\F_2$ coefficients.
Moreover, notice that only cells whose dimensions differ by 1 can have non-null value under the incidence function.

Both a simplicial complex with the simplicial incidence function and a critical set with the critical incidence functions belong to a class of combinatorial structures called Lefschetz complexes.
For Lefschetz complexes, we report only the basic notions and refer the reader to~\cite{Lefschetz1942} for more details about Lefschetz complexes or to ~\cite{Mrozek2009} for a notation closer to ours, where the Lefschetz complexes are called S-complexes.
\begin{deff}[Lefschetz complex]
A {\em Lefschetz complex} $(\cellular,\incfunc)$ {\em  over $\F_2$}  consists of a graded finite set $\cellular=\bigsqcup_{k\in\Z} \cellular_k$ along with an {\em incidence function} $\incfunc:\cellular\times \cellular\longrightarrow \F_2$ satisfying:
\begin{enumerate}[i)]
	\item \label{item:lef1} $\incfunc(\tau,\s)\neq 0$ implies that the dimensions of $\tau$ and $\s$ differ by 1
	\item \label{item:lef2} for every two cells $\tau^{k+2}$ and $\s^k$ in $\cellular$
then
\begin{equation*}
\sum_{\rho\in \cellular_{k-1}}\incfunc(\tau,\rho)\incfunc(\rho,\s) = 0
	\label{eq:ker in im for Lefschetz back}
\end{equation*}
\end{enumerate}
		\label{def:Lefschetz}
\end{deff}

A simplicial complex $\complex$ corresponds to a Lefschetz complex $(\cellular,\incfunc)$ with $\cellular$ graded by the simplex dimensions, that is  with $\cellular_k=\complex_k$, and  $\incfunc$ equal to the simplicial incidence function $\sif$.
Condition i) in Definition \ref{def:Lefschetz} is straightforward to check.
Condition ii) follows easily by noticing that only simplices $\rho$ satisfying $\tau>\rho>\s$ in $\complex$ lead to non null summands.
Moreover, in $\complex$, if existing, there are exactly two such $\rho$, thus summing up to zero in $\F_2$.

For a Lefschetz complex $(\cellular,\incfunc)$, in analogy with the simplicial case, if $\incfunc(\tau,\s)\neq 0$, we call $\tau$ a {\em cofacet} of $\s$ (resp. $\s$ a {\em facet} of $\tau$) and write $\tau>\s$ (resp. $\s <\tau$).
Moreover, each cell $\tau\in \cellular_k$ will be called a $k$-cell and often shortly denoted by $\tau^k$.

On the other hand, a critical set $\critical$ corresponds to a Lefschetz complex $(\cellular,\incfunc)$ by defining $\critical_k$ the set of all $k$-dimensional cells in $\critical$, by setting $\cellular_k=\critical_k$, and by choosing $\incfunc$ equal to the critical incidence function $\mif$.
We call $(\critical,\mif)$ a {\em critical Lefschetz complex over} $(\cellular,\incfunc)$.
We already noticed that Condition i) in Definition \ref{def:Lefschetz} is satisfied.
The fact that the function $\mif$ fulfills condition ii) follows from Theorem 8.10 in \cite{Forman1998} by applying the Remark \ref{rem:Lefschetz and chain complex condition equivalence} which we postpone.

\begin{deff}[Chain complex]
The {\em chain complex}  $\C(\cellular)=(\C_*(\cellular),\de_*)$ associated with the Lefschetz complex $(\cellular,\incfunc)$ consists of the family $\C_*(\cellular)=\{\C_k(\cellular)\}_{k\in\Z}$ of $\F_2$-vector spaces along with the collection of linear maps $\de_*=\{\de_k:\C_k(\cellular)\longrightarrow\C_{k-1}(\cellular)\}_{k\in\Z}$ defined as:
\begin{itemize}
	\item $\C_k(\cellular)$ is the vector spaces generated by $\cellular_k$ whose elements are called $k${\em -chains}
	\item $\de_k$ is called {\em boundary map} and defined by linear extension from the image of each cell $\tau^{k}$ defined by
\begin{equation}
\de_k(\tau)
=
\sum_{\s\in \cellular} \incfunc(\tau,\s)\s
	\label{eq:boundary map}
\end{equation}
\end{itemize}
\end{deff}

Notice that, by definition of Lefschetz complex, only $(k-1)$-cells can contribute to the image of the {\em boundary} $\de_k(\tau)$ for a $k$-cell $\tau$.
Moreover, condition (\ref{eq:ker in im for Lefschetz back}) in Definition \ref{def:Lefschetz} guarantees that $\im\de_{k+1}$ is included in $\ker\de_k$, that is
\begin{equation}
	\de_{k+1}\circ\de_{k+2} = 0
	\label{eq:delta quadro}
\end{equation}

\begin{rem}
	With our assumptions, condition (\ref{eq:delta quadro}) implies Condition \ref{item:lef2}) in Definition \ref{def:Lefschetz}.
	Indeed, by extending the terms in (\ref{eq:delta quadro}) when applied to a specific cell and manipulating the sums, we get that a sum of linearly independent elements gives the null vector. Hence, we deduce that all coefficients are null and coefficients corresponds the left hand side in Condition \ref{item:lef2}).
	\label{rem:Lefschetz and chain complex condition equivalence}
\end{rem}

By Remark \ref{rem:Lefschetz and chain complex condition equivalence}, we can look at Lefschetz complexes as a way of dealing with chain complexes in terms of their bases rather than the entire vector spaces.
Loops are formalized as $k$-chains with trivial boundary but such $k$-chains when bounding $(k+1)$-chains do not detect an actual hole.
A $k${\em-cycle} is an element in the kernel of $\de_k$.
A $k${\em-boundary} is an element in the image of $\de_{k+1}$.
The $\th{k}${\em-homology of the Lefschetz complex} $(\cellular,\incfunc)$ is defined as the quotient vector space of $k$-cycles over $k$-boundaries:
$$H_k(\cellular):=\bigslant{\ker\de_k}{\im\de_{k+1}}.$$

The combination of Theorems 7.3 and 8.2 in Forman's \cite{Forman1998} proves, in particular, the following
\begin{thm}[Homology invariance]
	Given a critical set $\critical$ over a simplicial complex $\simplicial$, the following isomorphisms hold
\begin{equation*}
	\forall k\in\Z, \quad
	H_k(\complex)\cong H_k(\critical)
	\label{eq:morse}
\end{equation*}
	\label{thm:homology preserving}
\end{thm}

\subsection{Multiparameter Persistent Homology}
\label{sec:PH}

Intuitively, persistent homology studies the homological changes along an increasing sequence of Lefschetz complexes called {\em filtration}.
We start by considering a finite total order $(I,\leq)$ and we call {\em grades} its elements.
In this paper, the grade set $I$ is to be thought of as a finite sampling in either $\R$ or $\Z$.
\begin{deff}[One-parameter filtration]
A filtration $\Fil \cellular$ of a Lefschetz complex $(\cellular,\incfunc)$ is a finite collection of Lefschetz subcomplexes $\Fil \cellular^u=(\cellular^u,\incfunc^u)$ in $\cellular$ indexed by $u\in I$ such that:
for all grades $u\leq v$, $\Fil \cellular^u$ is a Lefschetz subcomplex in $\Fil \cellular^v$
\end{deff}

\begin{deff}[Compatible discrete gradient]
Given a one-parameter filtration $\Fil \complex$ of a simplicial complex $\complex$ indexed on $I$, a discrete gradient $V$ over $\complex$ is called {\em compatible with } $\Fil \complex$ if, for each $(\s,\tau)\in V$ and any filtration grade $u\in I$, it holds that
\begin{equation*}
	\s\in \Fil\complex^u \quad\Rightarrow\quad \tau\in \Fil\complex^u
\end{equation*}
\label{def:compatible one-parameter}
\end{deff}

If a critical Lefschetz complex $(M,\mif)$ comes from a discrete gradient compatible with a filtration $\Fil \complex$, we call $(\critical,\mif)$ a {\em critical Lefschetz complex compatible with } $\Fil \complex$.

The latter concepts can be generalized to the multiparameter case. In place of a finite total order $(I,\leq)$, we can consider the partial ordered set $(I^n,\preceq)$ with $I^n$ the $n$-fold Cartesian product for some non-negative integer $n$ and, for any $u=(u_1,\dots,u_n),v=(v_1,\dots,v_n)\in I^n$,
\begin{equation*}
	u\preceq v \qquad \leftrightarrow \qquad
	\forall i\in\{ 1,\dots,n \}, u_i\leq v_i.
\end{equation*}
We call $(I^n,\preceq)$ the {\em grade poset} and, again, {\em grades} its elements.
If neither $u\preceq v$ or $v\preceq u$, we call the two grades $u$ and $v$ {\em incomparable} and {\em comparable} otherwise.
If $u\preceq v$ and $u\neq v$, we shortly wright $u\precneq v$.

\begin{deff}[Multiparamter filtration]
A multiparameter filtration, or simply {\em filtration} $\Fil \cellular$ of a Lefschetz complex $(\cellular,\incfunc)$ is a finite collection of Lefschetz subcomplexes $\Fil \cellular^u=(\cellular^u,\incfunc^u)$ in $\cellular$ indexed by $u\in I^n$ such that:
for all grades $u\preceq v$, $\Fil \cellular^u$ is a Lefscehtz subcomplex in $\Fil \cellular^v$
\end{deff}

In this paper, we are interested in a specific way of getting a filtration, that is by sublevel sets with respect to a function over $\cellular$.
A {\em filtering function} is a function $\phi:\cellular\longrightarrow I^n$ such that if $\s$ is a face of $\tau$, then $\phi(\s)\preceq \phi(\tau)$.
The {\em filtration induced by} $\phi$ is the family of Lefschetz subcomplexes $\Fil \cellular(\phi):=\{ \Fil \cellular(\phi)^u\}_{u\in I^n}$ in $(\cellular,\incfunc)$ with $\Fil \cellular(\phi)^u=(\cellular^u,\incfunc^u)$ defined by
\begin{align*}
	\cellular^u &:= \{ \s\in \cellular\ |\ \phi(\s)\preceq u \} \\
	\incfunc^u & := \incfunc \mbox{ restricted to } \cellular^u\times \cellular^u
\end{align*}

A subset $T$ is {\em closed} in $\cellular$ if, for all $\tau\in T$, the condition $\incfunc_\cellular(\tau,\s)\neq 0$ for some $\s \in \cellular$ implies $\s\in T$.
It is a known fact (Theorems 3.1, 3.2 in \cite{Mrozek2009}) that a closed subset $T$ in $\cellular$ is a Lefschetz subcomplex with incidence function taken by restriction.

\begin{rem}
 For each pair of grades $u\preceq v$, we have that $\cellular^u$ is closed in $\Fil \cellular^v$.
 Indeed, by definition of filtering function, faces cannot have higher grades than cofaces.
 This guarantees that $\Fil \cellular(\phi)$ is actually a filtration of $(\cellular,\incfunc)$.
\end{rem}

It is worth to remark that a one-parameter filtration $\Fil \cellular$ can always be thought of as a $\Fil \cellular(\phi)$ for some scalar-valued filtering function $\phi$.
On the contrary, if $n\geq 2$, filtrations induced by functions give a subclass of general filtrations.
For instance, filtrations of kind $\Fil \cellular(\phi)$ are {\em one-critical}, which according to \cite{Carlsson2009LNCS} , means that the minimal grade $u\in I^n$ a cell in $\cellular$ belong to in $\Fil \cellular(\phi)$ is unique.

%
Once we have a filtration, we can investigate how homology properties change from one step to another.
For any homology degree $k\in \Z$, we can associate each step $\Fil \cellular^u$ with its homology space $H_k(\Fil \cellular^u)$.
Moreover, since for all grades $u\preceq v$, the inclusion of corresponding Lefschetz complexes preserves cycles and boundaries, we get induced a linear map $\iot{u,v}:H_k(\Fil \cellular^u)\longrightarrow H_k(\Fil \cellular^v)$ at homology level, not necessarily injective since cycles can possibly become boundaries by adding cells.

The {\em persistent} $\th{k}${\em -homology } relative to the the grades $u\preceq v$ is the image of $\iota^{u,v}$ as a subspace in $H_k(\Fil \cellular^v)$, that is the space of all the homology classes of $H_k(\Fil \cellular^u)$ which are persistent in $H_k(\Fil \cellular^v)$.
The global information of persistent homology for all possible grades $u\preceq v$ is encoded in the {\em persistence module}.
\begin{deff}[Persistence module]
	The {\em persistence} $\th{k}${\em -module} $H_k(\Fil \cellular)$ of the filtered complex $\Fil \cellular$ consists of:
\begin{itemize}
	\item the collection of $\F_2$-vector spaces $H_k(\Fil \cellular^u)$, for each filtration step $u$
	\item the collection of all inclusion-induced linear maps $\iota^{u,v}:H_k(\Fil \cellular^u)\longrightarrow H_k(\Fil \cellular^v)$, for each pair of grades in $I^n$ satisfying $u \preceq v$.
\end{itemize}
\label{def:persistence module}
\end{deff}

In the case of $n=1$, we talk about {\em one-parameter persistent homology}, or simply one-parameter persistence.

Now, we are ready to formalize  the main property of the discrete gradient retrieved by the algorithm we are proposing in this paper.
\begin{deff}[Compatible discrete gradient]
Given a filtration $\Fil \complex$ of a simplicial complex $\complex$ indexed on $I^n$, a discrete gradient $V$ over $\complex$ is called {\em compatible with} $\Fil \complex$ if, for each $(\s,\tau)\in V$ and any filtration grade $u\in I^n$, it holds that
\begin{equation*}
	\s\in \complex^u \quad\Rightarrow\quad \tau\in \complex^u
\end{equation*}
\end{deff}

If a critical Lefschetz complex comes from a discrete gradient compatible with a filtration $\Fil \complex$, we call $(\critical,\mif)$ a {\em critical Lefschetz complex compatible with } $\Fil \complex$.
We already know from Theorem \ref{thm:homology preserving}  that the homology of a simplicial complex $\complex$ is preserved by any critical Lefschetz complex $(\critical,\mif)$ over $(\complex,\sif)$.
In fact, the filtration structure can be also preserved.
The following result generalizes to the multiparameter case Theorem 4.3 in \cite{mischaikow2013reductionPH} and equivalently Corollary 2 in~\cite{Allili2017dgci}.
\begin{thm}[Persistence module invariance]
	Given a filtration $\Fil \simplicial$ of a simplicial complex $(\simplicial,\sif)$ and a critical Lefschetz complex $(\complex,\mif)$ compatible with $\Fil \complex$, it holds that
\begin{equation}
	\forall k\in \Z, \quad
	H_k(\Fil \complex) \cong H_k(\Fil \critical)
	\label{eq:constrained morse}
\end{equation}
	\label{thm:correspondence pers module}
\end{thm}

This result guarantees that studying a critical Lefschetz complex compatible with a given filtration is equivalent to study the original filtered complex.
As an advantage, in the critical Lefschetz complex,  we have generally fewer cells to deal with.

\section{Related work}
\label{sec:related}

In this section we review the related work on the computation of persistent homology and multi-parameter persistent homology.

\subsection{Computing persistent homology}
\label{subsec:related-ph}
In the one-parameter case, computing the persistence module with coefficients in a field means reducing the {\em boundary matrix} via the standard algorithm~\cite{Edelsbrunner2002}. The latter algorithm has cubic complexity in the worst case. For this reason new approaches have been studied to improve efficiency. We present the resulting optimizations divided in three groups: {\em integrated}, {\em annotation-based}, and {\em preprocessing}.

{\em Integrated optimizations} aim at improving the efficiency of the standard approach by either reducing the number of steps reguired to get to a reduced matrix or by progressively removing columns during the computation. These approches exploit the total order defined on simplices of the simplicial complex to improve efficiency. We classify as integrated optimizations the {\em Twist} algorithm~\cite{Chen2011} the {\em Row} algorithm~\cite{DeSilva2011}, the approach based on sparsity presented in \cite{zigzag2011}, the one based on {\em Spectral sequences} \cite{edelsbrunner2008persistent}, and the {\em Chunk} algorithm~\cite{Bauer2014}.

{\em Annotation-based techniques}~\cite{Dey2014,Boissonat2013,Busaryev2012} take advantage of an efficient data structure, namely the annotation matrix, to efficiently compute the persistent co-homology of a complex.

{\em Preprocessing optimizations} aim at reducing the size of the input filtered complex while preserving the output persistence diagram. In \cite{Dlotko2014}, homology preserving techniques, such as reductions, coreductions \cite{mrozek2009coreduction,Mrozek2010coreductionPH,Dlotko2011} and acyclic subspaces \cite{Mrozek2008}, are adapted to the case of persistent homology. Approaches rooted in discrete Morse Theory \cite{Forman1998} compute a discrete gradient $V$ compatible with the input filtration. The theoretical results in \cite{mischaikow2013reductionPH} guarantees that the chain complex constructed from $V$ has the same persistence module of the input complex.
Many algorithms have been developed for computing a discrete gradient from a function sampled at the vertices of a cell complex. The algorithm described in \cite{King2005} is the first one to introduce a divide-and-conquer approach for computing a Forman gradient on real data. However, it has the main drawback of introducing many spurious critical simplices. Two approaches have been defined in \cite{Vijay12,Vijay12a} for 2D and 3D images respectively. Focusing on a parallel implementation, they provide a substantial speedup in computing the discrete gradient still creating spurious critical simplices. In \cite{Robins2011}, a dimension-agnostic algorithm is proposed that processes the lower star of each vertex independently. It has been proved that up to the 3D case, the critical cells identified are in one-to-one correspondence with the topological changes in the sublevel sets, i.e. no spurious critical simplices are created. An efficient implementation of \cite{Robins2011}, focused on regular grids, is discussed in \cite{Gunther2012} while, for simplicial complexes, the same algorithm as been extended to triangle \cite{Felle14} and tetrahedral meshes \cite{Weiss2013}. The first dimension independent implementation for simplicial complexes is presented in \cite{Fuga14}.\\


\subsection{Computing multiparameter persistent homology}
\label{subsec:related-mph}

The first difference we encounter when computing multi-parameter persistent homology is that we no longer have any complete descriptor for the persistence module \cite{carlsson2007multidimensional}. As a result, either we compute the full persistence module or we compute invariants that deliver only partial information about the multi-paramter persistent homology.
The first algorithm for the persistence module retrieval is proposed in \cite{Carlsson2009LNCS}, where the three tasks of computing the $k$-boundaries, $k$-cycles and their quotients at each multigrade $u$ are translated into {\em submodule membership problems} in computational commutative algebra. As drawbacks, the algorithm introduces an artefact dependency on the chosen basis and implies high computational costs in terms of time: $O(m^4n^3)$, where $m$ is the number of simplices in the complex and $n$ is the number of independent parameters in the multifiltration.
%
Another approach for computing the persistence module is proposed in \cite{Gafvert2016thesis}. The algorithm acts on the multifiltration at chain level rather than at homology level.
First, $k$-cycles and $k$-boundaries are expressed in terms of the same basis along the multifiltration at the chain level. Then, the Smith Normal Form reduction~\cite{Munkres1984, Agoston2005} is applied at each multigrade $u$ in the multifiltration leading to a worst time complexity of $O(m^3\bar{\mu}^n)$, where $m$ is the number of simplices, $n$ the number of parameters in the multifiltration, and $\bar{\mu}:=\max_{i=0,\dots,n}\mu_i$, with $\mu_i$ the number of multigrades in the multifiltration along the $\th{i}$-axis. The algorithm has been implemented in the Topcat library \cite{topcat} and distributed in public domain.
A non-complete descriptor for MPH is the {\em rank invariant}, introduced in \cite{carlsson2007multidimensional} for each pair of multigrades $u\preceq v$ as the rank of the corresponding inclusion-induced map, that is the number of homology classes from multigrade $u$ still persistent at multigrade $v$.
The rank invariant value over a single pair $(u,v)$ can be easily derived from the Topcat persistence module representation.
However, the full rank invariant computation requires the iteration of this simple procedure for all possible multigrades satisfying $u\preceq v$ which multiplies the complexity by $\frac{1}{2}\mu^2$, where $\mu$ is the cardinality of all multigrades considered in the multifiltration which is typically very large.

\paragraph{The persistence space} The {\em persistence space} \cite{Cerri2013dgci} is equivalent to the rank invariant but it practically enhances computational performances by avoiding to precompute the persistence module.

The persistence space can be computed based on the {\em foliation method} \cite{Biasotti2008JMIV}. With such approach the persistence space is constructed incrementally by slicing the space of the input multi-parameter filtrations and by constructing a number of one-parameter filtrations on which classic persistence homology is computed. The persistence pairs obtained on each slice form the persistence space. The first approach to the persistence space retrieval was limited to the case of $\th{0}$-homology  \cite{Biasotti2008JMIV}. Then, in \cite{Cagliari2010}, the foliation method is applied to higher homology degrees. An approximate version of the persistence space is proposed in \cite{Biasotti2011} for two-parameter filtrations, also called {\em bifiltrations}: a selection of slices is performed to guarantee a fixed tolerance for the matching distance~\cite{Cagliari2010} among persistence spaces. This method finds applications for shape comparison in the PHOG library \cite{Biasotti2013} where authors use the approximate persistence space to deal with photometric attributes.

Limitedly to bifiltrations, a complete representation of the persistence space is computed by the RIVET visualization tool \cite{Lesnick2015arXiv} which is available online at \url{http://rivet.online}. The approach uses the bigraded Betti numbers~\cite{Knudson2008,Eisenbud2005syzygies} to locate $\lambda_i$ multigrades, i.e., where changes in the homology of degree $i$ happen, with $i=1,2$. This procedure requires time $O(m^3\lambda)$, where $\lambda$ is the product of $\lambda_1\lambda_2$ and allows to idenfiy an arrangements of lines such that all filtrations along these lines have the same barcode template. A {\em barcode template} is constructed in $O(m^3\lambda + (m + \log \lambda)\lambda^2)$. The barcode template encodes the set of bars (i.e., persistence pairs) for every valid filtration in the space of bifiltrations. The actual length for each bar is computed on the fly, upon request, in linear time with respect to $m$.

\paragraph{Multiparameter optimization methods}

Most optimization methods developed for classic persistent homology have not yet found a counterpart in the multiparameter case. So far, the only approach that seem still feasible is simplifying the input filtration into a new one with less cells and less grades. Limited to the study of $\th{0}$-homology the algorithm proposed in \cite{Cerri2006} is the first approach capable of reducing the size of an input complex $\simplicial$ without affecting its persistence module.

The approach proposed in \cite{Allili2017JSC} can be seen as a Morse-based method generalizing to the multiparameter case the one proposed in \cite{King2005}. The algorithm computes a discreate gradient field having the same persistence module of the input complex. Like its one-parameter counterpart \cite{King2005}, it suffers from introducing many spurious critical simplices. In a successive paper \cite{Allili2017dgci} a new approach is introduced generalizing the idea of \cite{Robins2011} of constructing the discrete gradient locally inside the lower star of simplices of $\simplicial$. The resulting discrete gradient is proved to induce a Morse complex with the same persistence module, and then the same persistence space, as the original multifiltration. However, the algorithm requires a global ordering of all the simplices of $\simplicial$ and cannot be applied to real-world data. In Section \ref{sec:correctness}, we will further discuss this issue compared with our approach that can be seen as a divide-and-conquer generaliation of \cite{Allili2017dgci}.

\section{Locally computing a discrete gradient over multiparameters}
\label{sec:algo}

In this section we present our new algorithm for computing a discrete gradient vector field compatible with a multiparameter dataset. For ease of exposition, we describe the method by focusing on simplicial complexes though it is valid for any cell complex satisfying the intersection property such as {\em cubical complexes}.

In Section \ref{sec:outline} we provide a high-level description of the algorithm workflow. A detailed description of the auxiliary functions will be provided in Section \ref{sec:aux-func}, while in Section \ref{sec:complexity} we discuss the algorithm's complexity.

\subsection{Outline of the algorithm}
\label{sec:outline}

The proposed algorithm receives a multiparameter dataset in input and produces a compatible discrete gradient. In what follows, we describe the input multiparameter dataset as a pair $(\simplicial,f)$ where $\simplicial$ is a $d$-dimensional simplicial complex and $f:\simplicial_0\longrightarrow\R^n$ is a vector-valued function defined on the vertices of $\simplicial$. Without loss of generality we require the function $f$ to be component-wise injective. In applications, any function can be transformed into a component-wise injective one by means of simulation of simplicity \cite{Edelsbrunner1990}. The obtained output is a pair $(V,M)$ where $V$ is the set of paired simplices of $\simplicial$ and $M$ is the set of critical (unpaired) simplices completely defining the discrete gradient.

We recall that a discrete gradient is compatible to $f$ if for each pair of simplices $(\sOne,\sTwo)$ in $V$, $\sOne$ and $\sTwo$ have the same multigrade (see Section \ref{sec:background}). Then, the main objective of the algorithm is that of decomposing $\simplicial$ according to $f$ so to compute pairings between cells belonging to the same multigrade, possibly in parallel.\\

The algorithm consists of three main steps: vertex-based decomposition, multigrade grouping, and pairings computation. The main workflow is described in Algorithm \ref{alg:cdg}.

The first objective is that of decomposing $\simplicial$ to obtain a first rough subdivision of the simplices (line 2). In this step we only require that simplices belonging to the same multigrade also belong to the same group in the decomposition. This is achieved by the function \texttt{ComputeDiscreteGradient} as follows:

\begin{itemize}
	\item we compute an indexing $\indexing:\simplicial_0\longrightarrow\R$ for the vertices of $\simplicial$. The indexing is extended to the other simplices $\sOne\in\simplicial$ by setting $\indexing(\sOne):=\{ \indexing(v)\ |\ v\in\simplicial_0 \wedge v\ll\sOne \}$ and is required to be {\em well-extensible}, i.e., $\indexing$ satisfies, for all simplices $\sOne,\sTwo\in\simplicial$, the following property:
	\begin{equation}
		f(\sOne)\preceq f(\sTwo)
			\qquad\Rightarrow\qquad
			\indexing(\sOne)\leq \indexing (\sTwo).
	\label{def:well-extensible}
	\end{equation}

	\item We subdivide $\simplicial$ into lower stars according to $\indexing$. We recall the the star of a simplex $\sOne$ is defined as the set of cofaces of $\sOne$.  Then, we define the index-based lower star of a simplex $\sOne$ as the set of cofaces having value of $\indexing$ lower or equal to $\sOne$. Formally,
	\begin{equation*}
		\Low_\indexing(\sOne) :=
			\{ \sTwo\in\Star(\sOne) \ |\ \tilde{\indexing}(\sTwo) \leq
				\tilde{\indexing}(\sOne) \}.
	\label{eq:index-lower}
	\end{equation*}
\end{itemize}

The well-extensible indexing, in combination with the index-based lower star provide two fundamental properties for our algorithm:

\begin{itemize}
		\item each simplex $\sOne$ belongs to the indexed based lower star of exactly one vertex (Lemma \ref{lem:l inside low} in Section \ref{sec:correctness})
		\item if two simplices have the same multigrade, then they belong to the index-based lower star of the same vertex (Lemma \ref{lem:representative lowerstar} in Section \ref{sec:correctness}).
\end{itemize}

That is, a well-extensible indexing and the index-based lower stars implicitly provide a valid decomposition for the domain $\simplicial$. Thanks to the former properties we can guarantee that by processing the vertices independently we are not missing any valid pairing (see Section \ref{sec:correctness} for the formal proof).

The next step requires grouping the simplices of $\Low_I(v)$, with $v \in S_0$, having the same multigrade (lines 3-5). The latter is done by explicitly computing the index-based lower star for each vertex $v$ (function \texttt{ComputeIndexLowerStar}) and by subdividing the resulting set of simplices such that simplices with equal value of $\Filtering$ end up in the same set. This is done by the auxiliary function \texttt{SplitIndexLowerStar} which organizes the simplices and returns $K_v$, a list of sets where each set contains simplices with the same multigrade.

In the last step (lines 6-9), each multigrade $\Lset \in K_v$ is independently processed by the auxiliary function \he\ responsible for computing the actual pairings. Paired and critical simplices found in the multigrade set $\Lset$ will contribute to the final discrete gradient. Since simplices are subdivided based on their multigrade, each simplex $\simplicial$ appears in a exactly one level set and it will be classified, as either paired or critical, only once. This makes the approach embarrassingly parallel.

\begin{algorithm}[t!]
    \caption{ComputeDiscreteGradient($\simplicial,\filtering$)}
 \label{alg:cdg}
\begin{algorithmic}[1]
	\REQUIRE $\simplicial$ a simplicial complex
	\REQUIRE $\filtering:\simplicial_0\longrightarrow\R^n$ component-wise injective function
	\ENSURE $\gradient$ list of simplex pairs \LONGCOMMENT{discrete gradient compatible with $\Filtering$}
	\ENSURE $\critical$ list of simplices \LONGCOMMENT{critical cells of $\gradient$}
	\STATE $\gradient,\critical$ are empty lists
	\STATE $\indexing\leftarrow$ \texttt{ComputeIndexing}$(\simplicial_0,\filtering)$
	\LONGCOMMENT{$\indexing$ is a well extensible indexing with respect to $\filtering$}
	\FORALL[independently from the order] {$v$ in $\simplicial_0$}
	  \STATE $\Low_\indexing(v)\leftarrow$ \texttt{ComputeIndexLowerStar}$(v,\indexing,\simplicial)$\\
	  \STATE $K(v)\leftarrow$ \texttt{SplitIndexLowerStar}$(\filtering,\Low_\indexing(v))$\\
	  \FORALL[independently from the order]{${\Lset}$ in $K_v$}
	  \STATE $(\gradient_{{\Lset}},\critical_{{\Lset}})\leftarrow$ \texttt{HomotopyExpansion}$(\simplicial,{\Lset})$
	      \STATE append $\gradient_{{\Lset}}$ to $\gradient$
	      \STATE append $\critical_{{\Lset}}$ to $\critical$
	    \ENDFOR
	\ENDFOR
	\RETURN $(\gradient,\critical)$
\end{algorithmic}
\end{algorithm}

\subsection{Auxiliary functions}
\label{sec:aux-func}

This section provides additional information about the auxiliary functions we use Algorithm \ref{alg:cdg} following the order of appearance.

The first auxiliary function is \texttt{ComputeIndexing} which is used for computing a well-extensible indexing on the vertices of $\simplicial$. There are many ways to obtain a well-extensible indexing $\indexing$, we have chosen to sort all the vertices according to the values of the first component of $\filtering$. The total order obtained naturally generates an indexing which is guaranteed to be well-extensible as, for each pair of simplices $\sOne$ and $\sTwo$, $\Filtering(\sOne)\preceq \Filtering(\sTwo)$ implies $\Filtering_1(\sOne)\preceq \Filtering_1(\sTwo)$. Thus, a vertex $v\in\sTwo$ exists such that $\Filtering_i(v) \geq \Filtering_i(w)$ for every vertex $w\ll\sOne$. This implies $\indexing(v)\geq \indexing(w)$, for every vertex $w\ll\sOne$ and we conclude that $\tilde{\indexing}(\sOne)\leq \tilde{\indexing}(\sTwo)$.

Next, \textbf{\texttt{ComputeIndexLowerStar}} is used for computing the index-based lower star of a vertex from the indexing $\indexing$. The function extracts the set of simplices incident into a vertex $v$. We assume that each $k$-simplex $\sOne$ is represented by the list of its $k+1$ vertices $[v_0, v_1, \dots,v_k]$ stored in decreasing order of $\indexing$, i.e. $I(v_0) > I(v_1) > \cdots > I(v_k)$.

The computed index-based lower stars are then subdivided in independent sets by \texttt{SplitIndexLowerStar} according to multigrades. This function initializes an associative array mapping from a multigrade (a vector of floats) to the list of simplices sharing the same multigrade. We recall that the filtration values $\Filtering$ are assumed to be associated to the vertices of $\simplicial$ only. For any other simplex $\sOne$ the filtration value is computed for each component $i$ as $\Filtering_i(\sOne):=max_{v \in \sOne} \Filtering_i(v)$.


 \begin{algorithm}[t!]
     \caption{HomotopyExpansion($\cellular,\Lset$)}
 \label{alg:homexp}
\begin{algorithmic}[1]
        \medskip
  \REQUIRE $\cellular$, a simplicial complex,
  $\Lset$, a list of cells in $\cellular$ forming a level set w.r.t. $\Filtering$
 \ENSURE $\gradient_{\Lset}$ list of discrete vectors,
 $\critical_{\Lset}$ list of simplices
 \STATE set $\gradient_{\Lset},\critical_{\Lset}$ to be empty lists
 \STATE set \texttt{Ord0},\texttt{Ord1} to be empty ordered lists
 \STATE set \texttt{declared} to be an array of length $|\Lset|$ with  Boolean values equal to \texttt{false}\\
\FORALL{$\cTwo$ in $\Lset$}
   \IF{\texttt{num\_undeclared\_facets}($\cTwo,\Lset$)$=0$}
     \STATE insert $\cTwo$ into \texttt{Ord0}
   \ELSIF{\texttt{num\_undeclared\_facets}($\cTwo,\Lset$)$=1$}
     \STATE insert $\cTwo$ into \texttt{Ord1}
     \ENDIF
 \ENDFOR
 \WHILE{\texttt{Ord1}$\neq\emptyset$ or \texttt{Ord0}$\neq\emptyset$}
     \WHILE{\texttt{Ord1}$\neq\emptyset$}
       \STATE $\cTwo\leftarrow$ the first element in \texttt{Ord1}
\LONGCOMMENT{$\cTwo$ is removed from \texttt{Ord1}}
       \IF{\texttt{num\_undeclared\_facets}($\cTwo,\Lset$)$=0$}
 	       \STATE insert $\cTwo$ into \texttt{Ord1}
       \ELSE
           \STATE $\cThree\leftarrow$ \texttt{unpaired\_facet}($\cTwo,\Lset$)
\LONGCOMMENT{$\cThree$ is removed from \texttt{Ord0}}
 	        \STATE add $(\cThree,\cTwo)$ to $\gradient_{\Lset}$
 	        \STATE \texttt{declared}$[\cThree]$ , \texttt{declared}$[\cTwo]\leftarrow$\texttt{true}
           \STATE \texttt{add\_cofacets}($\cThree$,$\Lset$,\texttt{Ord1})
           \STATE \texttt{add\_cofacets}($\cTwo,\Lset$,\texttt{Ord1})
       \ENDIF
     \ENDWHILE
     \IF{\texttt{Ord0}$\neq\emptyset$}
       \STATE $\cTwo\leftarrow$ the first element in \texttt{Ord0}
\LONGCOMMENT{$\cTwo$ is removed from \texttt{Ord0}}
       \STATE append $\cTwo$ to $\critical_{\Lset}$
       \STATE \texttt{declared}$[\cTwo]\leftarrow$ \texttt{true}
       \STATE \texttt{add\_cofacets}($\cTwo,\Lset$,\texttt{Ord1})
     \ENDIF
   \ENDWHILE
 \RETURN $(\gradient_{\Lset},\critical_{\Lset})$
 \end{algorithmic}
 \end{algorithm}

As a last step, function \he\ classifies simplices with the same multigrade. We present its pseudocode in Algorithm~\ref{alg:homexp}. The execution has no conceptual differences from the one described in~\cite{Robins2011}. A $k$-simplex $\sOne$ and a $(k+1)$-simplex $\sTwo$ are considered {\em pairable} only when $\sOne$ is the only unclassified facet of $\sTwo$. So, the main objective of \he\ is that of pairing as many simplices as possible and to classify them as critical only when no pairable simplices are available.

Two ordered lists \texttt{Ord0} and \texttt{Ord1} are used to keep track of those simplices that have exactly zero unpaired facets or one unpaired facet, respectively. Intuitively, simplices in \texttt{Ord0} are candidates to be classified as critical or as tails of arrows in a discrete vector, since they have no face to be paired with, while simplices in \texttt{Ord1} are the candidate to be heads or arrows in a discrete vector. The two lists are initialized by cycling on the simplices in the input set (lines 4 to 8 of Algorithm~\ref{alg:homexp}). The auxiliary function \texttt{num\_undeclared\_facets}$()$ is used to count the number of unclassified facets for each simplex. Both lists \texttt{Ord0} and \texttt{Ord1} are ordered in such a way to have faces taking priority over cofaces.
The array \texttt{declared} keeps track of the simplices already classified (i.e., either paired or declared critical). At the beginning, all entries of \texttt{declared} are set to \texttt{false}.

Inside the two nested while loops (lines 9 and 10) is where simplices are classified. If \texttt{Ord1} is not empty, we extract the first simplex $\sTwo$ from it and we verify if the number of unclassified facets of $\sTwo$ has not changed (lines 12 and 13). Notice that the number of unpaired facets can only decrease. If this number is now zero (i.e., its facet has been classified), we add $\sTwo$ to \texttt{Ord0}. Otherwise we retrieve its unique unclassified facet $\sOne$ (line 15), we add ($\sOne$,$\sTwo$) to the set of pairs $\gradient_{\Lset}$, and we update the array \texttt{declared} accordingly. After classifying $\sOne$ and $\sTwo$, all their cofacets are visited and added to either \texttt{Ord1} or \texttt{Ord0}, if they have the necessary number of unclassified facets (lines 18 and 19).

When no pairable simplex is available (i.e., \texttt{Ord1} is empty) the first cell in \texttt{Ord0} is extracted and declared critical (lines 21 to 23). All its cofacets are processed and added to \texttt{Ord1} if it is the case. The algorithm stops when both lists are empty. In Proposition 4 in~\cite{Robins2011}, authors show that we exit the outer while loop when all cells have been classified.

\subsection{Complexity}
\label{sec:complexity}

In this section we discuss the computational complexity of \cdg\  and its auxiliary functions. To fix notation, the parameters involved in the analysis are expressed in terms of cardinality $|\cdot|$ of sets. We indicate with $\Star({\cOne})$ the star of a simplex $\cOne \in \simplicial$ and with $\Star$ the star with maximal cardinality in the simplicial complex $\simplicial$. Notice that, in a $d$-dimensional simplicial complex, $|\Star|$ is not bounded by a constant and is possibly as large as $|\simplicial|$. This is not the case for regular cell complexes like, for example, cubical complexes.

To simplify the analysis and the exposition we make a few assumptions:
\begin{itemize}

\item for each simplex $\sOne \in \simplicial$, we assume $\Star(\sOne)$ to be computed and stored off-line. If computed on the fly, $\Star(\sOne)$ would require $O(|\Star(\sOne)|)$ \cite{Canino11}.

\item the ordered lists \texttt{Ord0}, \texttt{Ord1} are implemented as self-balancing binary search trees. Inserting, or removing an element from the tree has a logarithmic cost in the list's size.

\item For each $k$-simplex $\sOne\in \simplicial$, $\Filtering(\sOne)$ can be retrieved in $O(k+1)$ by retrieving the filtration values of the vertices of $\sOne$. We will overestimate this by always considering the dimension $d$ of the simplicial complex $\simplicial$.
\end{itemize}

These assumptions are consistent with the implementation of \texttt{ComputeDiscreteGradient} used in our experimental evaluation (see Section \ref{sec:exp-allili}).

\subsubsection{Analysis of the auxiliary functions}
Here, we present the time and storage costs of the auxiliary functions introduced in Section~\ref{sec:aux-func}.

For creating the well-extensible indexing with \texttt{ComputeIndexing} we sort the vertices according to a single component of the input function. This requires $O(|\simplicial_0|\cdot\log |\simplicial_0|)$ time and $O(|\simplicial_0|)$ extra space for storing the new ordering.

The lower star of each vertex is then computed with \texttt{ComputeIndexLowerStar}. The lower star is extracted from the precomputed star $\Star(v)$ by selecting those simplices having $v$ as first vertex. This requires $O(|\Star(v)|)$ operations.



Once a lower star is extracted, the level sets are created by means of \texttt{SplitIndexLowerStar}. This requires retrieving the filtration value $\Filtering(\sOne)$ for each simplex $\sOne$ in the index-based lower star $\Low_\indexing(v)$ of a vertex $v$. Searching for the set of cells with a specific multigrade takes at most $O(\log |\Low_\indexing(v)|)$.
Then, the overall cost of \texttt{SplitIndexLowerStar} is $C_{LS} = O(|\Low_\indexing(v)|\cdot (d + \log|\Low_\indexing(v)|) )$.


For the last step, \texttt{HomotopyExpansion} classifies the cells in each level set. Preparing the two lists requires $O(|\Lset| \cdot |\log(|\Lset|)$ as for each simplex $\sOne$ in $|\Lset|$, \texttt{num\_undeclared\_facets} requires visiting its facets which number is limited from above by a constant factor. Inserting each simplex in the list takes $O(\log|\Lset|)$.

Within the two while loops, each simplex enters a list at most once and it is also classified once. Then, for each simplex $\sOne$:
\begin{itemize}
		\item retrieving its facets (\texttt{num\_undeclared\_facets} or \texttt{unpaired\_facets}) requires a constant number of operations,
		\item retrieving its cofacets (\texttt{add\_cofacets}) takes at most $O(|\Lset|)$ as the number of cofacets is not limited by any constant number,
		\item inserting the simplex in a list takes $O(\log|\Lset|)$.
\end{itemize}
Overall the contribution of \texttt{HomotopyExpansion} is $C_{HE} = |\Lset|(|\Lset| + 2\log(|\Lset|))$

\subsubsection{Analysis of \cdg\ algorithm}
\label{subsec:complexity-cdg}

By analizing the worst case complexity of the single auxiliary function we obtain a worst case complexity of

$$ O\left( |\simplicial_0|\log|\simplicial_0| + \sum_{v \in \simplicial} \left(|\Star(v)| + C_{LS} + \sum_{\Lset \subseteq \Low_\indexing(v)}C_{HE} \right) \right) $$

For the internal summation we can notice that in the worst case $|\Lset|$ is as big as the entire index-based lower star. Thus, we can overestimate $$\sum_{\Lset \subseteq \Low_\indexing(v)} C_{HE} = O(|\Low_\indexing(v)|(|\Low_\indexing(v)|+2\log(|\Low_\indexing(v)|)))$$.

We recall that each $k$-simplex appears in the star of its $k+1$ vertices. If we overestimate the dimension of each simplex $k$ with the dimension of the complex $d$, we can rewrite $\sum_{v \in \simplicial} |\Star(v)|$ as $|\simplicial|(d+1)$.

In a similar fashion, every simplex appears in exactly one index-based lower star. Thus, we can rewrite $\sum_{v \in \simplicial} (|\Low_\indexing(v)|(|\Low_\indexing(v)|+2\log(|\Low_\indexing(v)|)))$ as $|\simplicial|(|\Low_\indexing(v)|+2\log(|\Low_\indexing(v)|)$ and
$C_{LS}$ as $|\simplicial|(d+\log(|\Low_\indexing(v)|)$.

Moreover, we notice that in the worst case $\Low_\indexing(v)$ is as big as $\Star$. Based on this observation we can rewrite the overall complexity as

$$ O(|\simplicial_0|\log|\simplicial_0| + |\simplicial|(d + \log|\Star| + |\Star|)) $$

We should also mention that in applications we are often interested in filtrations defined on low dimensional complexes (i.e., with $d$=2,3). In such cases the number of simplices in each star becomes negligible leading us to a worst case complexity of $O(|\simplicial_0|\log|\simplicial_0| + |\simplicial|)$.

\section{Proof of correctness and comparisons}
\label{sec:correctness}
In this section, we provide a formal proof of correctness for algorithm \cdg. We formalize the correctness statement as follows:
\begin{quote}
	``The vector field returned by algorithm \cdg, with input the multifiltration $(\cellular,\filtering)$, is a discrete gradient field $\gradient$ and the corresponding Morse complex  $\critical$ is compatible with $(\cellular,\filtering)$.''
\end{quote}
The discrete gradient retrieved by algorithm \cdg\ is generated by the outputs of the auxiliary function \he\  wich is run over a single level set $\Lset$ in the index-based lower star $\Low_\indexing(v)$ of some vertex $v\in\simplicial_0$.
The strategy to prove correctness consists in showing equivalence to the algorithm \matching\ introduced in~\cite{Allili2017dgci}, and fully proved to be correct in~\cite{Allili2015arXiv}.
In order to do so, in Section~\ref{subsec:correctness-matching} we first review how algorithm \matching\ acts.
In Section~\ref{subsec:correctness-proof}, we prove the equivalence of \cdg\ and \matching.

\subsection{Globally computing a discrete gradient for multiparameters}
\label{subsec:correctness-matching}
In this section, we recall the procedure applied by algorithm \matching~\cite{Allili2015arXiv,Allili2017dgci} to retrieve the same object as our proposed algorithm \cdg\  introduced in Section~\ref{sec:algo}.

\paragraph{Input assumptions}
As for the case of \cdg, the \matching\  algorithm acts on a simplicial complex $\simplicial$ and a function $\filtering:\simplicial_0\longrightarrow\R^n$ required to be component-wise injective on vertexes and extended to higher dimensional simplices by function $\Filtering$ as defined in Section~\ref{sec:algo}.
The pair $(\simplicial,\Filtering)$ defines a multifiltration of $\simplicial$ obtained by sublevel sets.
Additionally, the \matching\ algorithm requires an {\em indexing} $\Indexing$ on $\simplicial$, i.e., an injective map $\Indexing:\simplicial\longrightarrow\R$.
The indexing has to be compatible both to the coface partial order $\ll$ among simplices and to the value ordering under $\Filtering$.
Explicitly, $\Indexing$ has to satisfy, the following property for every $\sOne\neq\sTwo\in\simplicial$,
\begin{equation*}
\sOne\ll\sTwo \mbox{ or } \Filtering(\sOne)\precneq \Filtering(\sTwo)	\quad\Rightarrow\quad	\Indexing(\sOne)<\Indexing(\sTwo).
\label{statement:good J simplicial}
\end{equation*}

\paragraph{Description of \matching}
The algorithm processes all simplices in $\simplicial$ in a for-cycle.
Simplices have to be processed according to increasing values of the indexing $\Indexing$.
This implies that, as opposed to the local algorithm of Section~\ref{sec:algo}, \matching\  cannot be broken into a parallel or distributed approach.

An auxiliary vector \texttt{classified}  of length $|\simplicial|$ with Boolean entries is initialized with all entries set to \texttt{false}.
For each simplex $\sOne$, the algorithm \matching\ checks whether $\sOne$ is classified.
An already classified simplex is not processed.
A non-classified simplex $\sOne$ is passed to an auxiliary function extracting the lower star of $\sOne$ with respect to $\filtering$
\begin{equation*}
	\Low_\filtering(\sOne) :=
		\{ \sTwo\in\Star(\sOne) \ |\ \Filtering(\sTwo)\preceq\Filtering(\sOne) \}.
\label{eq:lower-star}
\end{equation*}
%
To do so, the auxiliary function visits $\Star(\sOne)$ to select each simplex $\sTwo$ satisfying condition $\Filtering(\sTwo)\preceq \Filtering(\sOne)$.
Afterwards, an auxiliary function equivalent to the one in the algorithm \he\ (pseudocode reported in Algorithm~\ref{alg:homexp}) is run with input $(\Low_\filtering(\sOne),\Indexing)$.
We recall that algorithm \cdg, instead, calls \he\ with input $(\Lset,i(\lex))$, where $\Lset$ is a level set inside an index-based lower star $\Low_\indexing(v)$ for some vertex $v$, and $i(\lex)$ is the lexicographic order imposed on simplices by $\indexing$ by ordering the vertexes in decreasing order of values of $\indexing$.
Algorithm \he\  returns a pair of lists $(\gradient_{\Low_\filtering(\sOne)},\critical_{\Low_\filtering(\sOne)})$ and all entries in the auxiliary vector \texttt{classified} corresponding to the simplices in the two lists are set to \texttt{true}.
The global output is given by the independent contributions of all pairs $(\gradient_{\Low_\filtering(\sOne)},\critical_{\Low_\filtering(\sOne)})$.
We denote by $P$ the set of all simplices $\sOne\in\simplicial$ such that $\Low_\filtering(\sOne)$ is processed by \matching, also called primary simplices.

\paragraph{Correctness for the algorithm \matching}
It follows from Proposition 5 in~\cite{Robins2011} implying that, over each $\Low_\filtering(\sOne)$, the output is a valid (local) discrete gradient.
The fact that the union of all the independent discrete gradients returned by the auxiliary function \he\  forms a discrete gradient is given by Theorem 3.8 in~\cite{Allili2015arXiv}.
Finally, the compatibility with the input multifiltration $(\cellular,\filtering)$ is guaranteed by Theorem 3.7 in~\cite{Allili2015arXiv}.
In particular, Proposition 3.6 in~\cite{Allili2015arXiv} directly implies that lower stars $\Low_\filtering(\cOne)$ for $\cOne\in P$ form a partition of $\cellular$.

\paragraph{Summing up}
Both algorithms \cdg\ and \matching\  build their output discrete gradient by running \texttt{HomotopyExpansion} on a partition of the input complex $\cellular$:
\begin{itemize}
	\item \texttt{Matching} finds the discrete gradient independently over each lower star $\Low_\filtering(\cOne)$ with $\cOne\in P$,
	\item \texttt{ComputeDiscreteGradient} finds the discrete gradient independently over each level set $\Lset\in K_v$ with $v\in\cellular_0$.
\end{itemize}
In the next section, we show the two algorithms to be equivalent by proving that the two partitions are the same.

\subsection{Proof of equivalence}
\label{subsec:correctness-proof}
Under the notation of Section~\ref{subsec:correctness-matching}, in this section, we prove the equivalence between algorithms \cdg\ and \matching. The proof of all statements of this section is postponed to the appendix.
First, we show that the two algorithms apply the auxiliary function \he\ to the same partition of the input simplicial complex $\simplicial$.
Then, we show the desired equivalence.
In order for the partition into level sets $\Lset$ belonging to $\Low_\indexing(v)$ for some vertex $v$ to be coherent with the partition into $\Low_\filtering(\sOne)$'s, it is crucial that the indexing $\indexing$ is well-extensible as defined in (\ref{def:well-extensible}).
Then, the following statement holds:

\begin{lem}
  Let $\indexing$ be a well-extensible indexing with respect to $\filtering$.
  Then, for every
  $ \sOne\in \simplicial$ there is exactly one vertex $v\in \simplicial_0$ such that $\Low_{f}(\sOne)\subseteq\Low_I(v)$
  and $v\in\sOne$.\\
\label{lem:l inside low}
\end{lem}


The lemma above states that each filtration-based lower star $\Low_f(\sigma)$ is contained in exactly one index-based lower star $\Low_I(v)$, with $v$ a vertex of $\sigma$.
The following lemma states that each level set $\Lset \subseteq L_I(v)$ coincides with the maximal of the lower stars $\Low_f(\sigma)$  contained therein.

\begin{lem}
  Let $\simplicial$ be a simplicial complex, $f:\simplicial_0\longrightarrow\R^n$ a component-wise injective function
  and $I:\simplicial_0\longrightarrow\R$ a well-extensible indexing map with respect to $f$. Fix $\Lset\in K_v$ for some $v\in\simplicial_0$.
  Then, there exists a unique simplex $\sOne\in\simplicial$ such that $\Low_f(\sOne)=\Lset$ and, moreover, $\sOne\in \Lset$.\\
  \label{lem:representative lowerstar}
 \end{lem}

Lemma \ref{lem:l inside low} and Lemma \ref{lem:representative lowerstar} are used to prove that both algorithms apply \texttt{HomotopyExpansion} to the same portions of the domain.

\begin{lem}\label{lem:same HomotopyExpansion}
 Let $I:\simplicial\longrightarrow\R$ be a well-extensible indexing with respect to $f$
 and $J:\simplicial\longrightarrow\R$ a suitable input indexing for \texttt{Matching}.
 Let $P$ be the set $\{ \sOne\in\simplicial\ |\ $\texttt{Matching} runs \texttt{HomotopyExpansion} over $\Low_f(\sOne)\ \}$.
 Then, for any level set $\Lset\in K_v$, there exists a $\sOne\in\simplicial$ such that
 $$
 \Lset= \Low_f(\sOne)
 \Leftrightarrow
 \sOne\in P
 $$
\end{lem}


Hence, the equivalence of the two approaches can be shown by focusing on the reduction of one input into the other.

\begin{prop}\label{prop:cdg to matching}
For every input $(\cellular,\filtering)$ for \texttt{ComputeDiscreteGradient}, there exists a suitable input indexing $\Indexing:\cellular\longrightarrow\R$ for \texttt{Matching} such that the output of \texttt{Matching}$(\cellular,\Filtering,\Indexing)$ equals that of \texttt{ComputeDiscreteGradient}$(\cellular,\filtering)$.\\
\end{prop}

As a corollary, we get the correctness of \texttt{ComputeDiscreteGradient}.
\begin{cor}
	Algorithm \texttt{ComputeDiscreteGradient} with input $(\cellular,\filtering)$ returns a discrete gradient $\gradient$  compatible with the multifiltration induced by $(\simplicial,\filtering)$.
\label{cor:correctness}
\end{cor}

The following proposition completes the equivalence between the algorithms \texttt{ComputeDiscreteGradient} and \texttt{Matching}.
In particular, it states that all possible gradients retrieved by the global algorithm \texttt{Matching} can be obtained by the local strategy of \texttt{ComputeDiscreteGradient}.
\begin{prop}\label{prop:matching to cdg}
 Let $(\cellular,\Filtering,\Indexing)$ be a suitable input for \texttt{Matching}.
  Let $\filtering:\cellular_0\longrightarrow\R^n$ be the restriction of $\Filtering$ to the vertexes.
  Then, the output of \texttt{Matching}($\cellular,\Filtering,\Indexing$) equals that of \texttt{ComputeDiscreteGradient}($\cellular,\filtering$), provided that, for each level set $\Lset$ under $\Filtering$,
 the function \texttt{HomotopyExpansion} is given $(\cellular,\Lset,\Indexing)$ as input.
\end{prop}


This last statement provides not simply correctness, but it also states that \texttt{ComputeDiscreteGradient} is as general as \texttt{Matching}.
This means that the introduction of the well-extensible indexing over the vertexes which is needed in algorithm \texttt{ComputeDiscreteGradient} can be performed for every possible input.

\subsection{Comparison of Asymptotical Complexity}
\label{subsec:complexity-comparison}

In this section, we compare the computational complexity of algorithm \cdg\ , provided in Section~\ref{sec:complexity}, with that of algorithm \matching\ as discussed in~\cite{Allili2017dgci}.
As reviewed in Section~\ref{sec:correctness}, \texttt{Matching} and \texttt{ComputeDiscreteGradient} apply \texttt{HomotopyExpansion} to exactly the same level sets. In \cite{Allili2017dgci}, authors express this cost as $O(|\simplicial|\cdot|\Star|\cdot\log|\Star|)$.
%
The actual difference in complexity between the two algorithms is relative to the different number of stars which need to be visited in order to apply \texttt{HomotopyExpansion}.
Indeed, \texttt{Matching} computes a lower star, for each level set of $\Filtering$.
Instead, in \cdg, computes a lower star for each vertex in the input complex.
This means that, in our case, the exact number of cells visited by \texttt{ComputeIndexLowerStar} is $|S|$.
Instead, the number of cells visited by \matching\  depends on the number of level sets $|P|$ in the input dataset, satisfying
$$
|\cellular_0|\leq |P| \leq |\cellular|.
$$
In the case where $|P|=|\cellular_0|$, the two algorithms visit the same number of cells.
In the case where $|P|=|\cellular|$, it means that each cell belongs to a different level set.
Since each $j$-cell has $\binom{j+1}{k+1}$ different $k$-dimensional faces, each $j$-cell belongs to exactly $\binom{j+1}{k+1}$ cell stars.
In that case, the amount of visited cells is given by
\begin{equation}
	\sum_{\cOne\in\cellular}
	|\Star(\cOne)| =
	\sum_{j=0}^d
	|\cellular_j|
	\sum_{k=0}^{j+1}
	 \binom{j+1}{k+1}.
\label{eq:star simplicial sum on cells}
\end{equation}

In Section \ref{subsubsec:experiments-allili-results} we provide experimental results showing that, even if both algorithms works linearly in the number of cells, our approach guarantees an improved scalability.


\subsection{Reducing a multiparameter filtration in practice}
\label{sec:exp-allili}

In this section we will compare experimentally our local preprocessing approach and the global matching algorithm introduced in~\cite{Allili2017dgci}. Each {\em original dataset} is formed by a pair $(\simplicial,\filtering)$, where $\simplicial$ is a simplicial complex and $\filtering:\simplicial\longrightarrow\R^n$ is a component-wise injective function. Here, we focus on the case where $\simplicial$ is a triangle mesh embedded in the Euclidean 3D space and $\filtering$ is a bifiltration that assign to each vertex its x and y coordinates (i.e., for $v=(x,y,z)$, $\filtering(v)=(x,y)$).
In order to guarantee a fair comparison, both algorithms have been implemented by using the FG\_Multi library \cite{mdg} which provides an efficient encoding for the triangle mesh as well as for the computed discrete gradient.

\paragraph{Representing a simplicial complex}
A triangle mesh $\simplicial$ is a simplicial $2$-complex formed by vertices, edges, and triangles. The FG\_Multi library \cite{mdg} implements an incidence-based data structure for compactly encoding the relations among these simplices.
Vertices and triangles are the only simplices which are explicitly encoded for a total of $|\simplicial_0|+|\simplicial_2|$ entities. Each vertex encodes the list of triangles incident while each triangle encodes a reference to its three vertices. Notice that, for each triangle $\sigma$ referencing a vertex $v$, we also have that $v$ references $\sigma$.  Then, if each triangle references three vertices the triangle-vertex relation costs $3|\simplicial_0|$ while the vertex-triangle relation doubles this cost leading to a total of $7|\cellular_2|+|\cellular_0|$. The filtering function $\filtering$ is stored for each vertex by encoding a vector of floating point values, one value for each parameter.

\paragraph{Discrete gradient representation}
The discrete gradient $\gradient$ is here encoded by adopting the representation described in \cite{Felle14}. The latter focuses on encoding all the gradient pairs locally to each triangle. The encoding uses the following rationale. Since each triangle $\sOne$ can be paired with at most three edges and each edge can be paired with two vertices, locally for each triangle we have 9 possible pairs. If we consider also the possible pairs between an edge and an adjacent triangle we get 12 possible gradient pairs and thus $2^{12}=4096$ possible combinations. However, a discrete gradient imposes certain restrictions, i.e. that each simplex can be involved in at most one pairing. As a consequence we have only 97 valid cases for a triangle. These cases can be encoded using only 1 byte per triangle and, thus, encoding the gradient only requires $|\cellular_2|$ bytes. Notice that the latter approach has been generalized to tetrahedral meshes \cite{Weiss2013} and $d$-dimensional simplicial complexes \cite{Fugacci2018gmod}.

\subsubsection{Experimental results}
\label{subsubsec:experiments-allili-results}

The dataset used in this comparison is originated by three triangle meshes. For each mesh we obtain two refined versions of the latter by recursively applying the Catmull-Clark algorithm~\cite{Catmull1978} to it.
The nine triangle meshes composing the final dataset are described in Table \ref{tab:datasets}. Column {\em Original} indicates the number of simplices composing the mesh. Column {\em Critical} indicates the number of unpaired (critical) simplices identified by both reduction approaches while the resulting compression factor is reported in column {\em Original/Critical}.
%

\begin{table}[h!]
\centering
\resizebox{.55\columnwidth}{!}{%
\begin{tabular}{l|c|rr|r}
	& & \multicolumn{2}{c|}{Cells}
	& Compression factor \\
	Dataset & Parameters & Original & Critical & Original/Critical  \\ 
\hline
\cellcolor{Gray}
& \cellcolor{Gray}
& 1.3M \cellcolor{Gray}
& 0.035M \cellcolor{Gray}
& 37.9 \cellcolor{Gray}
\\
\cellcolor{Gray}
& \cellcolor{Gray}
& 5.3M \cellcolor{Gray}
& 0.11M \cellcolor{Gray}
& 45.3 \cellcolor{Gray}
\\
\multirow{-3}{*}{Torus} \cellcolor{Gray} 
& \multirow{-3}{*}{2} \cellcolor{Gray}
& 21.5M \cellcolor{Gray}
& 0.77M \cellcolor{Gray}
& 27.7 \cellcolor{Gray}
\\ 
&
& 2.9M  
& 0.28M 
& 10.2 
\\
&
& 11.7M 
& 0.11M 
& 10.2 
\\
\multirow{-3}{*}{Sphere} 
& \multirow{-3}{*}{2}
& 47.1M 
& 0.46M 
& 10.1 
\\ 
\cellcolor{Gray}
& \cellcolor{Gray}
& 3.8M  \cellcolor{Gray}
& 0.4M \cellcolor{Gray}
& 9.5 \cellcolor{Gray}
\\
\cellcolor{Gray}
& \cellcolor{Gray}
& 15.2M \cellcolor{Gray}
& 1.6M  \cellcolor{Gray}
& 9.4 \cellcolor{Gray}
\\
\multirow{-3}{*}{Gorilla} \cellcolor{Gray} 
& \multirow{-3}{*}{2} \cellcolor{Gray}
& 60.9M \cellcolor{Gray}
& 6.4M  \cellcolor{Gray}
& 9.4 \cellcolor{Gray}
\\ 
\end{tabular}
}
\caption{\label{tab:datasets} Datasets used for the experiments. For each of them, we indicate the number of independent parameters in the multifiltration (column {\em Parameters}), the number of simplices in the original dataset (column {\em Original}), number of critical simplices retrieved by \texttt{ComputeDiscreteGradient} and \texttt{Macthing} (column {\em Critical}) and the compression factor (column {\em Original/Critical}).}
\end{table}

The experiments have been performed on a dual Intel Xeon E5-2630 v4 CPU at 2.20 Ghz with 64GB of RAM.
\begin{figure}[h]
\centering	\includegraphics[width=0.5\linewidth]{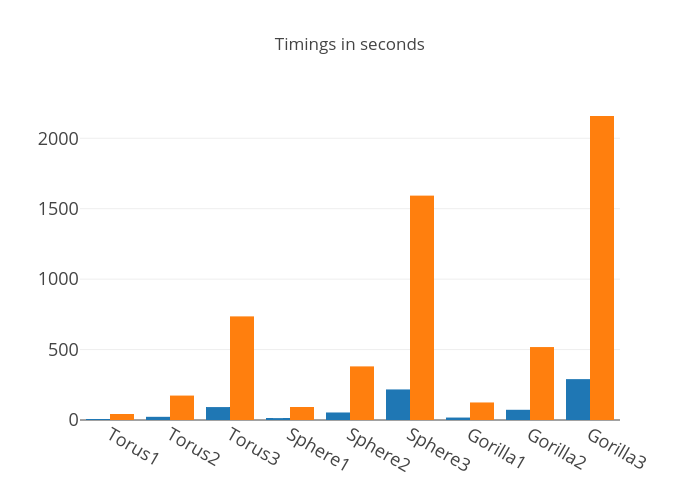}
\caption{\label{fig:TimevsAllili} Timings required by \texttt{ComputeDiscreteGradient} (in blue) and \texttt{Matching} (in orange).}
\end{figure}


Timings are shown in Figure \ref{fig:TimevsAllili}. The local approach takes between 0.89 seconds and 4.8 minutes to finish depending on the dataset and it is generally 7 times faster than our implementation of the global approach.
Time performances show the practical efficiency of the local approach compared to the global one.
As seen in \ref{sec:complexity}, the expected asymptotical complexity over a triangle mesh $\simplicial$ (i.e., $d=2$) is $O(|\simplicial|\log|\Star|)$ for both algorithms. For the datasets considered in these tests the number of simplices withing each star is negligible which makes the algorithm linear in the number of simplices.

\begin{figure}[h!]
\centering

    \includegraphics[width=0.5\linewidth]{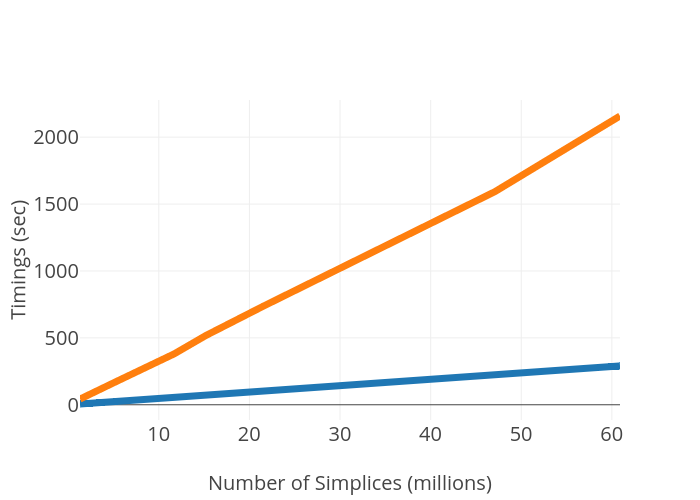}

\caption{\label{fig:linear} Trend in the timings for \texttt{ComputeDiscreteGradient} (in blue) and \texttt{Matching} (orange).}
\end{figure}

In Figure~\ref{fig:linear}, we show the trends as the number of simplices increases. This confirms our argument that the number of stars to be retrieved and visited has a direct consequence on the algorithm complexity.


%
In the local case we need to process each vertex star only while the global approach requires processing a star for each multigrade.

\begin{figure}[h!]
\centering

    \includegraphics[width=0.5\linewidth]{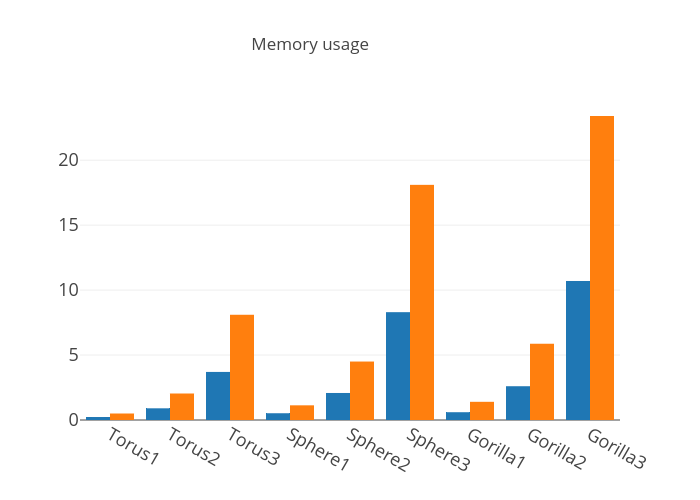}

\caption{\label{fig:MemvsAllili} Maximum peaks of memory (in gigabytes) required by \texttt{ComputeDiscreteGradient} (in blue) and \texttt{Matching} (in orange).}
\end{figure}


Other than time efficiency, our divide-and-conquer strategy also requires a limited use of memory. The memory consumption is shown in Figure \ref{fig:MemvsAllili} where we are reporting the maximum peak of memory used by the two algorithms. The local approach uses up to 10 gigabytes of memory versus more than 20 of the global approach.
The storage cost difference between the two implementations grows linearly in the number of simplices in the datasets.
As seen in \Cref{subsec:complexity-comparison}, the advantage is due to the different memory consumption at runtime. In particular, the global approach stores a global queue over all the simplices in the dataset and needs to track all classified cells. Both these steps are performed locally, over each level set, by the local approach.

\section{Computing multiparameter persistence homology on reduced datasets}
\label{sec:exp}

In this section, we evaluate the impact of our preprocessing method for the computation of the persistence module (Section \ref{sec:exp-topcat}) and of the persistence space (Section \ref{sec:exp-foliations}). Before presenting our results, we describe how to extract, from a discrete gradient $V$, the Morse complex that will be used as input of persistence computations.

\paragraph{Computing the Morse complex} We recall that a discrete gradient $V$ implicitly represents a Morse complex $M$ having the cells in one to one correspondence with the critical simplices of $V$. The incidence relations among the cells of $M$ are described by the gradient paths originating and having destination in a pair of criticial simplices. Intuitively, two cells of $M$ are incident to each other if there is a gradient path that connects the corresponding critical simplices in $V$.

To compute such relations we process all the critical cells of $V$. For each critical cell $\sOne$ of dimension $k$, a breadth-first traversal is performed as follows.
From $\sOne$ we extract its incident $(k-1)$-simplices. For each $(k-1)$-simplex we extract its paired simplex $\sOne_1$, if any. We apply the same rationale to $\sOne_1$ to continue the visit. As soon as we encounter a $(k-1)$-simplex $\sTwo$ which is unpaired (critical) we register the two cells $\sOne$ and $\sTwo$ as incident to each other.

In the worst case, computing incidences for a single critical simplex requires visiting all $k$-simplices multiple times. In particular $O(|\simplicial_k|^2)$, where $|S_k|$ is the numbere of $k$-simplices in $S$. Having a number of critical simplices of the same order of $|\simplicial_k|$ would bring the total worst-case complexity to be cubical in the number of simplices. In real cases, the extraction of the Morse complex is very efficient as each $k$-simplex belongs to a very limited set of gradient paths, possibly zero.

\subsection{Computing the persistence module}
\label{sec:exp-topcat}

In this section, we evaluate the impact of the reduction method to the computation of the persistence module. The persistence module will be computed by means of the open-source library Topcat~\cite{topcat}, which is currently the only available library for this task. Due to its strong limitations in terms of time and memory costs, we used a simplified dataset for our experiments. We use six triangle meshes of limited size, three representing a torus and the other three representing a sphere. The bifiltration used for each triangulation is defined by the x and y coordinates of the vertexes. Table \ref{tab:vstopcat} presents a description of the dataset. Number of entities (column {\em Cells}) and number of multigrades (column {\em Grades}) are reported for each simplicial complex (column {\em Original}) and each Morse complex (column {\em Reduced}).
%
\begin{table}[h!]
\centering
\resizebox{\columnwidth}{!}{%
\begin{tabular}{l | r r r r | r r r r | r }
       & \multicolumn{4}{c|}{Original} & \multicolumn{4}{c|}{Reduced} & Reduction Time \\
       &&& \multicolumn{2}{c|}{Persistence Module}
       &&& \multicolumn{2}{c|}{Persistence Module}\\
	Dataset & Cells & Grades & Time & Memory & Cells          & Grades & Time & Memory \\
	\hline
	\cellcolor{Gray}
	& \cellcolor{Gray}
	38 \cellcolor{Gray}
	& \cellcolor{Gray}
	8x8 \cellcolor{Gray}
	& 0.3 \cellcolor{Gray}
	& 0.24  \cellcolor{Gray}
	& 4  \cellcolor{Gray}
	& 5x5 \cellcolor{Gray}
	& 0.18 \cellcolor{Gray}
	& 0.1 \cellcolor{Gray}
	& 0.0264\cellcolor{Gray}
	\\
	\cellcolor{Gray}
    & \cellcolor{Gray}
    242 \cellcolor{Gray}
    & 42x42 \cellcolor{Gray}
    & 4.4 \cellcolor{Gray}
    & 0.86 \cellcolor{Gray}
    & 20 \cellcolor{Gray}
    & 10x10 \cellcolor{Gray}
    & 0.28 \cellcolor{Gray}
    & 0.2 \cellcolor{Gray}
    & 0.0244 \cellcolor{Gray}
    \\
    \multirow{-3}{*}{Sphere} \cellcolor{Gray}
    & 2882 \cellcolor{Gray}
    & 482x482 \cellcolor{Gray}
    & - \cellcolor{Gray}
    & -  \cellcolor{Gray}
    & 278 \cellcolor{Gray}
    & 92x89 \cellcolor{Gray}
    & 24.3 \cellcolor{Gray}
    & 1.5 \cellcolor{Gray}
    & 0.0473 \cellcolor{Gray}
    \\
    \multirow{3}{*}{Torus}
    & 96
    & 16x16
    & 0.5
    & 0.1
    & 8
    & 9x9
    & 0.25
    & 0.2
    & 0.0255
    \\
    & 4608
    & 768x768
    & -
    & -
    & 128
    & 65x66
    & 7.96
    & 2.4
    & 0.0643
    \\
    & 7200
    & 1200x1200
    & -
    & -
    & 156
    & 70x80
    & 12.05
    & 3.0
    & 0.0815
    \\
\end{tabular}
}
\caption{\label{tab:vstopcat} Timings (in seconds) and storage costs (in gigabytes) for the persistence module retrieval over the original (columns {\em Original}) and the corresponding reduced (columns {\em Reduced}) datasets. Columns {\em Cells} and {\em Grades} reports the number of cell in the dataset and the number of grades along each parameter, respectively. Missing entries indicate where the Topcat library run out of memory. Column {\em Reduction Time} explicit the timings (in seconds) for obtaining the reduced cell complex.}
\end{table}


%
The Topcat library uses the boundary matrices of the complex to compute the persistence module. Since it was designed to accept only multifiltrations defined on simplicial complexes we have modified the library to make it read multifiltrations defined over general cell complexes.

We compute the persistence module of each dataset, both the original and the reduced ones, and we measure time and storage consumption of the Topcat library. We report the results obtained in Table \ref{tab:vstopcat}, columns {\em Time} and {\em Memory}. These represent the timings (in seconds) and the memory (in Gigabyte) required for computing the persistence module.
Where no result is reported, the Topcat library runs out of memory.
We notice that timings are always in favor of the Morse complex. Where a comparison is possible, computing the persistence module on the Morse complex takes approximately half of the time than computing it on the original simplicial complex.

The memory consumption is the main bottleneck of the Topcat algorithm as it is mainly affected by the number of cells and the number of multigrades. The use of the reduction approach reduces this problem by shrinking the number of boths. In our experiment all successful executions have used a limited amount of memory, significantly below the machine limit of 64GB. This suggests a dramatic increase of memory usage in the ones where the computations have failed. For instance, the failure of the test over, for instance, the Sphere dataset with 2882 cells and 482x482 multifiltration multigrades suggests that computing the persistence module on a reduced dataset of the same size would fail as well. We should stress the fact that the objective of our experiment is that of evaluating the gain in performances when using our reduction approach and not that of overcoming the limitation of Topcat.
Column {\em Reduction Time} reports the partial timings required for computing the reduced cell complex. These include the timings contribution of running \cdg\ together with the retrieval of the boundary matrix through the algorithm~\cite{Fugacci2018gmod}.
We can notice that the measured reduction timings, ranging from 0.0244 to 0.0815 seconds are negligible with respect to the time required for computing the persistence module.

\subsection{Computing the persistence space}
\label{sec:exp-foliations}
In this section, we evaluate the impact of the reduction method on the computation of the persistence space. We recall that the persistence space can be computed via the {\em foliation method} introduced in~\cite{Biasotti2008JMIV}. The foliation method consists in restricting each multifiltration to several linearization, i.e.,  single-parameter filtrations, called {\em slices}. On each slice, any computational technique from single-parameter persistence can be applied to obtain a persistence diagram. The collection of persistence diagrams, gives an approximation of the persistence space. The number of slices to consider varies based on the application. As a rule of thumb, the more slices we consider, the more accurate is the approximation of the persistence space obtained.

\paragraph{The foliation method} The foliation method can be seen as a two-step approach for the computation of the persistence space. For sake of simplicity we present a description for the foliation method specific for a bifiltration $\phi$.

The first step consists of uniformely selecting $\omega^2$  lines of non-negative slope in the Euclidean plane.
First we compute the extremal values for the bifiltration $\phi$. For each component $i=1,2$, we compute parameters $C_i:=\max_{x\in\simplicial}\phi_i(x)$ and $c_i:=\min_{x\in\simplicial}\phi_i(x)$.
Each line $l$ that we will extract is determined by two parameters: $\lambda$, i.e. the slope coefficient, and $b$, i.e. the base point. For creating $\omega^2$ we uniformely select $\omega$ values for both $\lambda$ and $b$. Values of $\lambda$ range from 0 to $\frac{\pi}{2}$. Value of $b$ are computed as follow.
For each value of $\lambda$, we select the bisector of the II and IV quadrant with slope $\lambda$. The projections of points $(c_1,C_2)$ and $(C_1,c_2)$ over the bisector will limit the interval on which sampling the values of $b$.
All possible values for $\lambda$ and $b$ are combined to represent the $\omega^2$ possible lines.
Each line $l=(m,b)$ is the line of unit vector with $m=(\cos(\lambda),\sin(\lambda))$ and passing through $b$.

For each line extracted $l=(m,b)$ we create a new 1-dimensional filtration over the simplices of $\simplicial$. Each simplex $\sOne$ obtains the filtration value $\Phi^l$ according to $l$ as:
$\Phi^l(\sOne):=\min_{i=1,2}{m_i} \cdot \max_{i=1,2}\frac{\phi_i(\sOne)-b_i}{m_i}$. The obtained filtration is used to compute classic persitent homology. The resulting persistence pairs within each persistence diagram will form the approximated persistence space.\\

\subsubsection{Computing the persistence space of the Morse complex.}

In this subsection we present results for evaluating the impact of our reduction approach when computing the persistence space. The foliation method requires the choice of two parameters: the number of slices and the method used for computing classic persistent homology. In the following we will present results providing insights on both, either by varying the number of slices (between 2 and 100) or by varying the method for computing persistent homology.

Datasets considered are from the Princeton Shape Benchmark \cite{Shilane2004}. Table \ref{tab:100slices} describes the dataset and the corresponding results obtained when computing the persistence space by using 100 slices and by using the standard algorithm implemented in PHAT.
For each dataset reported in Table \ref{tab:100slices} the first row reports data regarding the original mesh while the second row describes the corresponding Morse complex computed by using our reduction method. For each input complex we show the number of cells (column {\em Cells}) and the average number of persistence pairs found per slice (column {\em Pairs}).

Timings are reported separately for the computation of the Morse complex (column {\em Reduction}), for the extraction of slices (column {\em Line Extraction}) and for the actual computationa of the persistence space (column {\em Foliations Time}). The latter is formerly subdivided into three partial timings accounting for the construction of the boundary matrix (column {\em Building Pers. input}), computation of persistent homology (column {\em Computing Persistence}), reindexing of the persistence pairs according to the multifiltration (column {\em Reindexing Pers. output}). Column {\em Foliations Total} shows the sum of the partial timings.

\begin{table}[h!]
\centering
  \resizebox{\columnwidth}{!}{%
\begin{tabular}{l | c c c | c | c | c c c c }
       & & & & & & \multicolumn{4}{c}{Foliations Time}
       \\
     &  &  & & Reduction & Line  & Building & Computing & Reindexing & Foliations
    \\
	Dataset & Cells & \multicolumn{2}{c|}{Pairs} & Time & Extraction  & Pers. input & Persistence & Pers. output         & Total
    \\ \hline
\cellcolor{Gray}
& 9491 \cellcolor{Gray}
& 4744 \cellcolor{Gray}
	& \cellcolor{Gray}
& \cellcolor{Gray}
& \cellcolor{Gray}
& 9.04 \cellcolor{Gray}
& 1.91 \cellcolor{Gray}
& 1.45 \cellcolor{Gray}
& 12.42 \cellcolor{Gray}
\\
\multirow{-2}{*}{Shark} \cellcolor{Gray}
& 1111 \cellcolor{Gray}
& 554	\cellcolor{Gray}
	& \multirow{-2}{*}{\cellcolor{Gray}(81.4)}
& \multirow{-2}{*}{0.11} \cellcolor{Gray}
& \multirow{-2}{*}{0.86} \cellcolor{Gray}
& 1.15 \cellcolor{Gray}
& 0.21 \cellcolor{Gray}
& 0.84 \cellcolor{Gray}
& 2.22 \cellcolor{Gray}
 \\
& 10861
& 5426
& \multirow{2}{*}{(8.8)}
& \multirow{2}{*}{0.12}
& \multirow{2}{*}{0.63}
& 10.21
& 2.11
& 1.53
& 13.87
\\
    \multirow{-2}{*}{Turtle}
& 1197
& 594
&
&
&
& 1.22
& 0.22
& 0.84
& 2.29
\\
\cellcolor{Gray}
& 27826 \cellcolor{Gray}
& 13873 \cellcolor{Gray}
&  \cellcolor{Gray}
&  \cellcolor{Gray}
&  \cellcolor{Gray}
& 27.49 \cellcolor{Gray}
& 5.65  \cellcolor{Gray}
& 2.69 \cellcolor{Gray}
& 35.85 \cellcolor{Gray}
\\
\multirow{-2}{*}{Gun} \cellcolor{Gray}
& 3144 \cellcolor{Gray}
& 1532 \cellcolor{Gray}
& \multirow{-2}{*}{(10.2)}\cellcolor{Gray}
& \multirow{-2}{*}{0.28} \cellcolor{Gray}
& \multirow{-2}{*}{0.65}\cellcolor{Gray}
& 3.18 \cellcolor{Gray}
& 0.60 \cellcolor{Gray}
& 0.99 \cellcolor{Gray}
& 4.77 \cellcolor{Gray}
 \\
& 119081
& 59349
& \multirow{2}{*}{(79.5)}
& \multirow{2}{*}{1.14}
& \multirow{2}{*}{0.85}
& 118.14
& 26.56
& 10.33
& 155.91
\\
\multirow{-2}{*}{Piano}
& 10955
& 5286
&
&
&
& 11.09
& 2.26
& 1.65
& 15.01
 \\

\end{tabular}
}
\caption{\label{tab:100slices} Timings (in seconds) required for computing the persistence pairs on 100 uniformly sampled slices.
Datasets are reported by rows.
For each triangle mesh, the first row is for the original dataset and the second one for the reduced dataset considered over the same 100 slices.
Column {\em Cells} reports the number of cells in the multifiltration.
Column {\em Pairs} reports the average number of persistence pairs found per slice. In parantheses, the number of pairs with positive persistence (equal for original and reduced datasets).
Reported timings are subdivided into phases a) (column {\em Reduction Time}), b) (column {\em Line Extraction}), and c) (column {\em Foliations Time}).
The latter subdivided into step 1) (column {\em Building Pers. input}), step 2) (column {\em Computing Persistence}), and 3) (column {\em Reindexing Pers. output}).}
\end{table}


We notice that, by reducing the number of cells of approximately one order, we get a one-order reduction on all timings.
Looking at column {\em Line Extraction} we notice that the extraction of the lines has little to no differences across the triangle meshes. This happens because in this case we are always considering the same number of slices. For the partial timings, the highest contribution is shown in column {\em Building Pers. Input}. This is the part where the cells are sorted by increasing values under $\Phi^l$ and reindexed according to this values. Both this phase and the following one (i.e., the actual computation of persitent homology) are affected by the number of input cells, indeed the results for the reduced dataset reflect the one-order reduction in the number of cells.
Results shown in column {\em Reducing Pers. Output} depend on the number of persistence pairs found. The difference in the results obtained with the original triangle mesh and the corresponding Morse complex suggests that our reduction step let us consider fewer spurious persistence pairs.
Column {\em Foliation Total} indicates timings for computing the persistence space as a whole. The total timings required by a reduced dataset range from a minimum of 2.22 seconds (Shark triangle mesh) to a maximum of 15.01 seconds (Piano triangle mesh), whereas, the original datasets require from 12.42 (Shark triangle mesh) to 155.91 seconds (Piano triangle mesh).

\begin{figure*}[t]
	\centering
	\begin{tabular}{c}
		\includegraphics[width=0.9\linewidth]{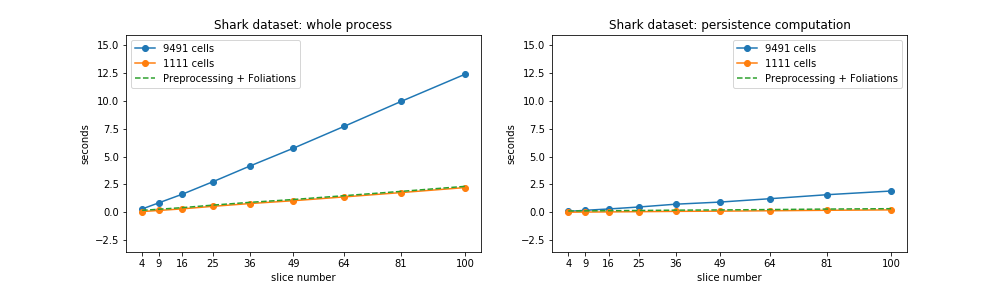} \\
		(a)	\\\includegraphics[width=0.9\linewidth]{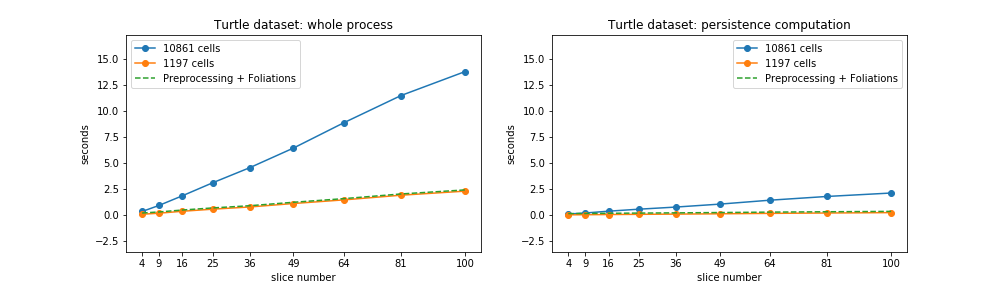} \\
		(b) \\
        \includegraphics[width=0.9\linewidth]{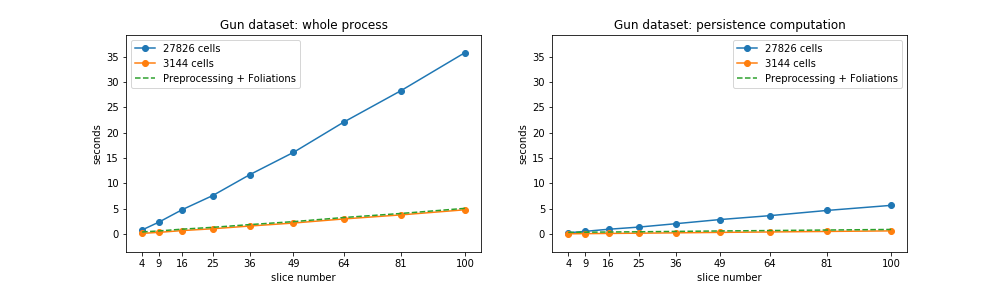} \\
		(c)	\\\includegraphics[width=0.9\linewidth]{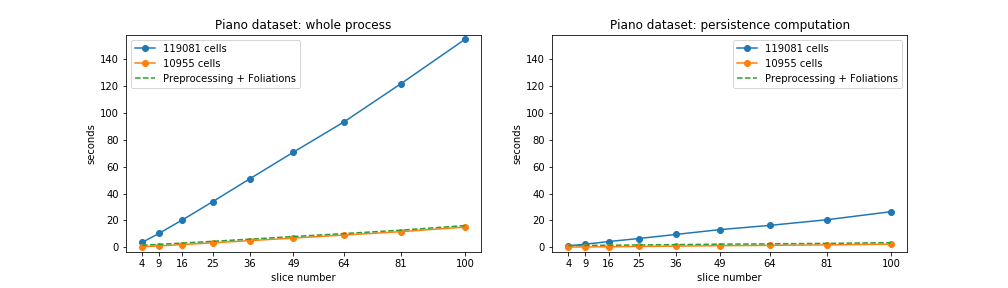} \\
        (d) \\
	\end{tabular}
	\caption{Time performances plotted with respect to a number of slices varying form 4 to 100 over the same dataset. Datasets considered are triangle meshes:
    (a) Shark,
    (b) Turtle,
    (c) Gun, and
    (d) Piano.
In all the figures, on the left,
performances are indicated in blue for the original dataset and in orange for the corresponding reduced dataset.
On the right, we show the same plotting with respect to step 2) in the foliation phase only.
}
	\label{fig:foliations-over-slices}
\end{figure*}


\clearpage

\paragraph{Varying the number of slices}

In Figures~\ref{fig:foliations-over-slices}, we compare the time performances achieved by the foliation method using a number of slices ranging from 4 to 100. The one-parameter persistence over each slice is computed by the standard algorithm implemented in PHAT. For each dataset, we show, on the left, the global timings for the foliation phase and, on the right, the partial timings required by the computation of persistent homology.

Blue lines indicate results obtained for the triangle meshes, the green dotted line presents results obtained with the Morse complexes accounting for both the reduction algorithm and the foliation step. Orange lines indicate results obtained with the Morse complexes exclusively for the foliation phase. As we can see, orange and green lines almost overlapp indicating that the preprocessing step used for computing the Morse complex is almost negligible with respect to the computation of the persistence space.

We also notice the linear dependency of the process from the number of slices. For reduced datasets (orange line), the slope coefficient is smaller than for the original datasets (blue line). This is more evident for global timings suggesting that a preprocessing reduction is preferable independently from the number of considered slices. Notice that, limitedly to the computation of persistent homology when using 4 slices, we get the blue line just below the green dashed line. This is the only exception where the preprocessing step could be avoided.

Our tests confirm that the time complexity in the foliation method primarily depends on the number of slices considered.
Our reduction approach impacts on the performances by simply reducing the number of cells to be processed.
Moreover, our tests show that the proposed preprocessing is effective also for a small number of slices.

\paragraph{Varying the persistent homology computation algorithm}


\begin{figure}[p]
	\centering
	\begin{tabular}{c c}
		\includegraphics[width=0.45\linewidth]{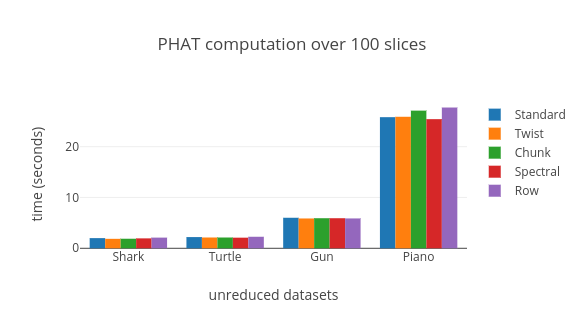} &
		\includegraphics[width=0.45\linewidth]{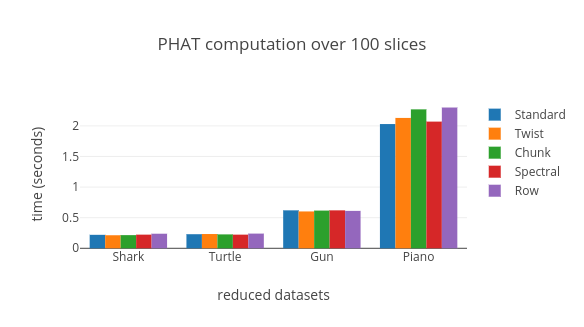} \\
		(a) & (b) \\
	\end{tabular}
	\caption{Timings for 100 slice computations via the five algorithms for persistence implemented in the PHAT library.
Original datasets (a) are compared to reduced datasets (b).
    }
	\label{fig:phatMethods}
\end{figure}


In Figure \ref{fig:phatMethods}, we report the results obtained by using the five algorithms implemented in PHAT for computing persistent homology on the original (a) and reduced datasets (b) over 100 slices. Also here we can notice that computing persistent homology on the reduced datasets takes an order of magnitude less than on the original triangle mesh.

In our test, performances of the standard algorithm are comparable to the other appraoches implementing optimizations. This was not expected according to~\cite{Bauer2014}.
This can be explained, in part, by the low dimension of the chosen meshes and their limited size, but we should also notice that in the foliation method we have to run the same algorithm multiple times. For this reason, the number of slices may have a more profound impact on the overall timing than the optimization implemented on the single slice. On top of that, our results already suggest that a multiparameter reduction strategy is preferable since it can be computed only once and used for all the slices.

\section{Concluding remarks}
\label{sec:conclusions}

In Section~\ref{sec:algo}, we have proposed a new preprocessing algorithm for MPH suitable for applications to real-sized data sets.
We have highlighted the local character of our approach as opposed to the global character of the equivalent existing approach in~\cite{Allili2017dgci}.
Our complexity analysis makes it clear that the two preprocessing algorithm might have the same worst-case time complexity depending on the input.
In fact, we have discussed how the presence of multiparameter in place of one parameter affects the average case rather than the worst-case time.

Concerning 	 the issue of quantifying the advantage of our proposed MPH preprocessing to computing the persistence module,
our local MPH preprocessing increases of up to about 50 times the size of the input complex that can be treated, and up to about 250 times the size of the filtration that can be treated.
In all considered datasets (rather small), the reduction allows to complete the pipeline. Some non-reduced datasets have failed for running out of memory.
These failures for rather small datasets suggest that, at the moment, our preprocessing is not enough to make the persistence module computation feasible over real-size data.
In particular, we detected memory costs as a bottleneck for current persistence module computational methods.
Optimizations of current algorithms require better handling of memory usage in terms of size of the multifiltration and number of cells in the input complex.

Concerning 	 the issue of quantifying the advantage of our proposed MPH preprocessing to computing the persistence space,
our local MPH preprocessing shows its advantages in all considered datasets.
The foliation method allows to retrieve the persistence space by multiple iterations of PH computations.
One goal was that of evaluating the tradeoff between the number of iterations and the advantages of the MPH preprocessing.
We found that, in all considered datasets, the reduced datasets outperforms the corresponding original dataset, regardless of the number of iterations applied.
Instead, when limited to the PH computation timings, only the case of 4 slices shows advantages for non-processed datasets. This is coherent with other comparisons made for PH efficiency such as~\cite{Bauer2014,Otter2017} (non-processed datasets should be preferable for few iterations).
Moreover, we have found that our MPH preprocessing is preferable over all considered PH optimized algorithm to be iterated.
Finally, we notice that, in our test, performances of the standard algorithm are comparable to the considered optimizations.
This was not expected according to~\cite{Bauer2014}.
Our choice of triangle meshes datasets, that is with low geometric dimensions and cell stars limited in size, may explain that results.

\subsection{Future Work}
\label{subsec:future}
The results discussed in this paper suggest future works in multiple directions.
From a computational point of view, the results obtained motivate the need for studying and developing finer implementations of the available techniques, especially in the case of the persistence module retrieval.

Additionally, we think that the idea of a discrete gradient compatible with a multifiltration deserves further insights from the theoretical point of view.
Currently, we are working on defining a notion of optimal reduction for a multifiltration. The optimality should extend the property satisfied by the algorithm~\cite{Robins2011} equivalent to our proposed one in the case of a single parameter filtration.

Moreover, comparisons between reductions based on critical cells of a discrete gradient and  other bifiltration reductions, such as the one implemented in RIVET~\cite{Lesnick2015arXiv}, should be studied both theoretically and computationally.
We are working on this problem by trying to relate critical cells in a multifiltration to the notion of {\em multigraded Betti numbers}.

Finally, the study of critical cells of a multifiltration may be addressed from the topology-based visualization perspective.
The critical cells of a multifiltration might be interpreted as a fully discrete counterpart to other approaches to visualize mutual behavior of multiple scalar fields, such as Pareto sets~\cite{Huettenberger2014} or Jacobi sets~\cite{Edelsbrunner2004}.
At the moment, we are working on defining an incidence structure among critical cells arranged into a compact graph to be compared to other similar structures available for piecewise linear functions such as the Reachability graph~\cite{Huettenberger2014}.

\section{Acknowledgements}

This work has been partially supported by the US National Science Foundation [grant number IIS-1116747].
The authors wish to thank Michael Kerber for interesting discussion on the results.

\section*{References}
\bibliography{ext_bib}

\section*{Appendix}
\label{sec:appendix}
In this appendix, we report the proofs of results we omitted in~\Cref{subsec:correctness-proof}.

\paragraph{Proof of Lemma 1}
Let $\sOne$ be a simplex in $\simplicial$.
It is easy to see that the $\Low_\indexing(v)$'s form a partition of $\simplicial$.
Hence, there exists a unique vertex $v\ll\sOne$ such that $\sOne$ belongs to $\Low_I(v)$.
Let $\sTwo$ be a simplex in $\Low_{\tilde{f}}(\sOne)$. By definition of lower star, $\sTwo\gg\sOne$ and $\tilde{f}(\sTwo)\preceq \tilde{f}(\sOne)$.
The former condition implies that $\tilde{I}(\sTwo)\geq \tilde{I}(\sOne)$ and that $v\in\sTwo$.
The latter condition together with the assumption on $I$ being well-extensible give $\tilde{I}(\sTwo)\leq \tilde{I}(\sOne)$.
Hence, $\tilde{I}(\sTwo)=\tilde{I}(\sOne)=I(v)$, which concludes the proof.

\paragraph{Proof of Lemma 2}
Let $v$ and $\Lset$ be as in the lemma statement.
In order to prove uniqueness, suppose there are two simplices $\sOne,\sOne'\in\simplicial$ such that $\Low_f(\sOne)=\Lset=\Low_f(\sOne')$.
   Notice that, $\sOne,\sOne'\in \Lset$, since any simplex belongs to its own lower star.
   Moreover, condition $\Low_f(\sOne')=\Low_f(\sOne)$ implies that $\sOne'\in\Low_f(\sOne)$ and $\sOne\in\Low_f(\sOne')$ at the same time.
   By definition of lower star, we get in particular $\sOne'\ll\sOne$ and $\sOne\ll\sOne'$, that is $\sOne'=\sOne$.
   In order to prove existence, we define $\sOne$ to be the intersection of all simplices belonging to $\Lset$.
   We know that $v\in\sTwo$ for any $\sTwo\in \Lset$ and that when two simplices intersect they do it in a single shared face, so $\sOne$ is a non-empty simplex.
   Notice that $\sOne$ belongs to $\Low_I(v)$ and, since $f$ is component-wise injective, for any $i=1,\dots,n$ $\sTwo\in \Lset$, it holds that $\tilde{f}_i(\sTwo)=f_i(w)$ with $w\in\sTwo$.
   This implies, $w\in\sOne$ and, thus, $\sOne\in \Lset$.
   We claim that $\Low_f(\sOne)=\Lset$.
   Indeed, $\Lset$ is trivially part of $\Low_f(\sOne)$ since any $\sTwo\in \Lset$ has the same value under $\tilde{f}$ and $\sTwo\gg\sOne$.
   Conversely,, let $\sTwo$ be a simplex in $\Low_f(\sOne)$.
   Since $I$ is well-extensible, the Lemma \ref{lem:l inside low} ensures that $\sTwo\in\Low_I(v)$.
   Notice that, for a general $\sTwo\in\Low_f(\sOne)$, it holds that $\tilde{f}(\sTwo)=\tilde{f}(\sOne)$.
   Being $\sOne\in \Lset$, we get $\sTwo\in \Lset$, which proves our claim and concludes the proof.

\end{document}